\def\year{2022}\relax
\documentclass[letterpaper]{article} 
\usepackage{aaai22}  
\usepackage{times}  
\usepackage{helvet}  
\usepackage{courier}  
\usepackage[hyphens]{url}  
\usepackage{graphicx} 
\usepackage{natbib}  
\usepackage{caption} 
\usepackage{algorithm}
\usepackage{algpseudocode}
\usepackage{amsmath}
\usepackage{amsfonts}
\usepackage{amssymb}
\usepackage{graphicx} 
\usepackage{subcaption}
\usepackage{xcolor}
\usepackage{afterpage}
\usepackage{enumitem}
\usepackage{url}

\newcommand\blankpage{%
    \null
    \thispagestyle{empty}%
    \addtocounter{page}{-1}%
    \newpage}

\DeclareCaptionStyle{ruled}{labelfont=normalfont,labelsep=colon,strut=off} 
\usepackage{newfloat}
\usepackage[switch]{lineno}
\usepackage{listings}
\floatstyle{ruled}
\newfloat{listing}{tb}{lst}{}
\floatname{listing}{Listing}

\title{A Dynamic Meta-Learning Model for Time-Sensitive \\ Cold-Start Recommendations}
\author{Krishna Prasad Neupane, Ervine Zheng, Yu Kong, Qi Yu}
\affiliations{
 Rochester Institute of Technology \\
 
 \{kpn3569, mxz5733, yu.kong, qi.yu\}@rit.edu
 }

 \pdfoutput=1
\begin{document}
\maketitle

\begin{abstract}
We present a novel dynamic recommendation model that focuses on users who have interactions in the past but turn relatively inactive recently.  Making effective recommendations to these {\em time-sensitive cold-start users} is critical to maintain the user base of a recommender system. Due to the sparse recent interactions, it is challenging to capture these users' current preferences precisely. Solely relying on their historical interactions may also lead to outdated recommendations misaligned with their recent interests. The proposed model leverages historical and current user-item interactions and dynamically factorizes a user's (latent) preference into time-specific and time-evolving representations that jointly affect user behaviors. These latent factors further interact with an optimized item embedding to achieve accurate and timely recommendations. Experiments over real-world data help demonstrate the effectiveness of the proposed time-sensitive cold-start recommendation model.

\end{abstract}

\section{Introduction}
Recommender system has long been used as an effective means to improve user experience and to provide personalized recommendations in diverse fields such as media, entertainment, and e-commerce. \cite{sun2014collaborative,xie2018factorization}. One effective way of recommendation is via Collaborative Filtering (CF) \cite{goldberg1992using,su2009survey}, which recommends items based on similar users' preferences. CF assumes that users who had similar interactions with some items in the past are likely to have the same preference on other items, and leverages the observed user-item interactions to make predictions for the missing parts, which indicate the potential items of interest to users. Matrix Factorization (MF) is one commonly used CF technique that exploits user and item latent factors to capture their inherent attributes. Most existing MF methods model user preferences and item attributes as static factors \cite{koren2008factorization,koren2009matrix}. Some recent efforts consider the dynamic changes in the user-item interactions by modeling user preferences as shifting latent factors over time, allowing them to provide more timely recommendations \cite{dpf,gultekin2014collaborative,sun2014collaborative}.

When user-item interactions are very sparse, making accurate recommendations based on limited information is highly challenging, which is usually referred to as the cold-start problem. A viable solution to handle the cold-start problem is to utilize extra side information such as item descriptions, user profiles \cite{uyangoda2018user}, and social relationships \cite{yao2014dual}. While various side information is widely available for items (i.e., movies, songs, and books), the user-related side information is typically very scarce as acquiring users’ personalized information may be practically difficult due to privacy issues. Several recent works adopted meta-learning for few-shot learning in the recommender systems to alleviate the cold-start problem \cite{vartak2017meta,lee2019melu,du2019sequential}. Meta-learning aims to learn global knowledge from the historical information of many similar tasks (users) and then provide quick adaptation for a new task (user) with limited interaction information.
Although the meta-learning approaches show promising results \cite{vartak2017meta,lee2019melu}, they are primarily designed for static settings and hence not effective in providing {\em timely recommendations that best reflect users' current interests}.

\begin{figure}
\begin{subfigure}{.24\textwidth}
  \centering
  \includegraphics[width=0.98\linewidth]{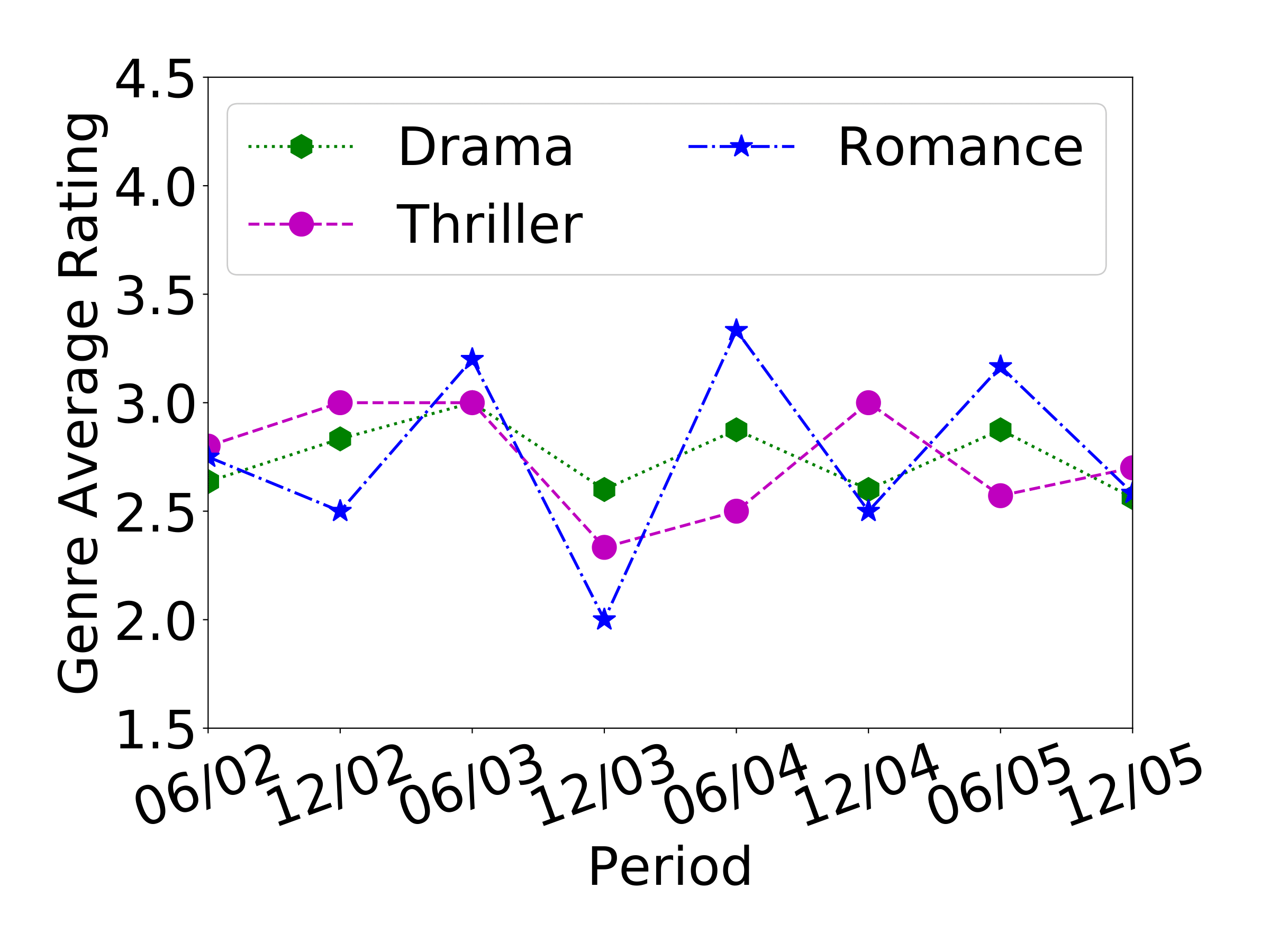}
  \vspace{-2mm}
  \caption{}
  \label{time_spec}
\end{subfigure}%
\begin{subfigure}{0.24\textwidth}
\vspace{-3mm}
  \centering
  \includegraphics[width=0.98\linewidth]{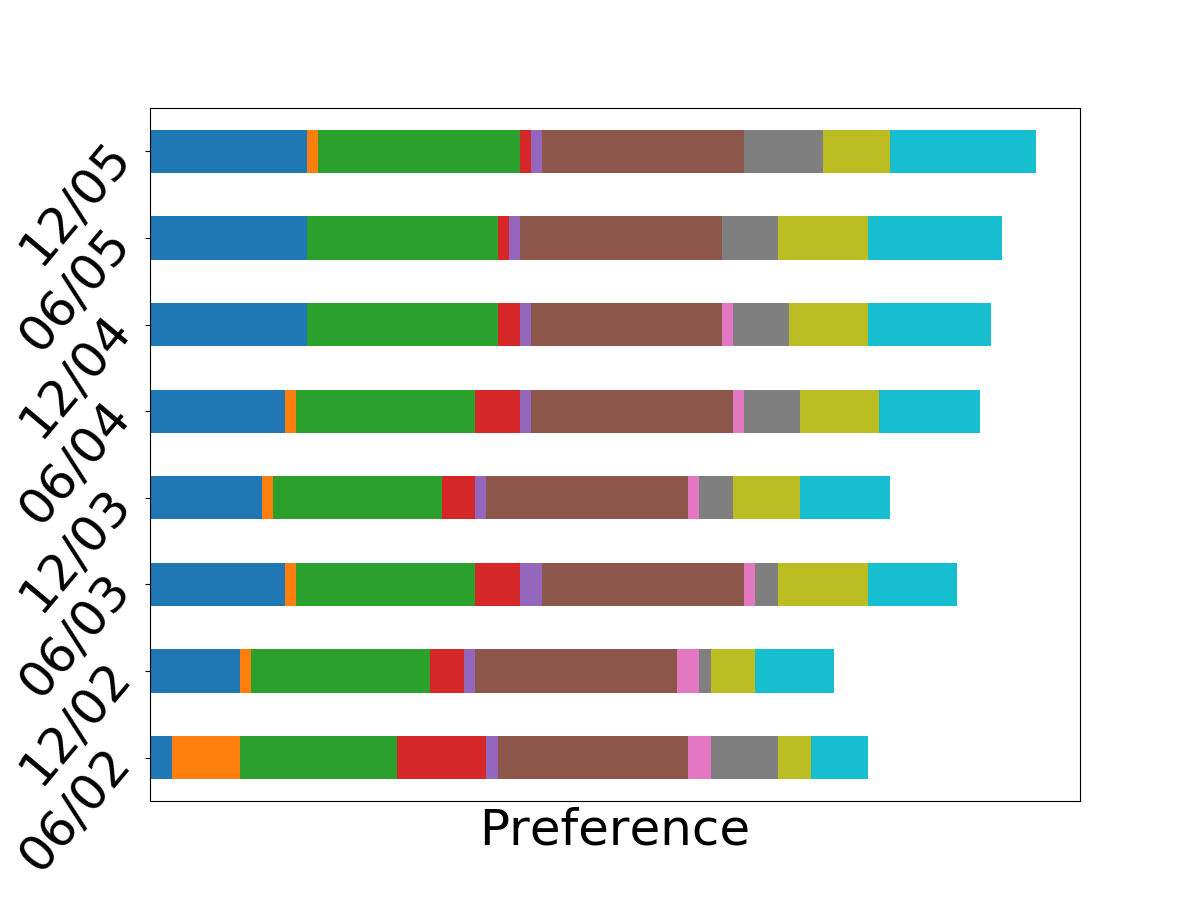}
  \vspace{-1mm}
  \caption{}
  \label{time_evol}
\end{subfigure}
\vspace{-2mm}
\caption{For User (ID:2181970): (a) shows dynamic changes of ratings for three genre types over time (b) demonstrates how user's latent preferences evolve over time.
}
\label{fig_1}
\vspace{-4mm}
\end{figure}

Figure~\ref{time_spec} shows that the average movie ratings for three genre types of an example user from the Netflix dataset, which oscillate significantly from 2002 to 2005. The average rating indicates an overall satisfaction on each type of genre items that the user interacts with for each period, and serves as an indicator for  the change of users' preference.
Those changes are usually caused by the interplay of two contributing factors. First, a user's preference over different types of items (e.g., movie or music) may change over time, which we refer to as the {\em time-evolving factor}. Second, a user's behavior in a specific period may vary significantly from other periods due to the impact of money/time budget or other external causes, which we refer to as the {\em time-specific factor}. Given sufficient user interactions over time, both factors could be effectively learned to provide accurate and timely recommendations.
However, in practice, many users may have interactions in prior periods but become largely inactive with few interactions in recent periods for various reasons. We define these users as the \textit{time-sensitive cold-start users} due to scarce recent interactions. Solely relying on the historical interactions of these users may lead to outdated recommendations that do not match their recent interests. Furthermore, the limited recent interactions pose a user cold-start problem for the current period that makes existing CF-based techniques less effective.

The main challenge with these time-sensitive cold-start users is to simultaneously capture their most recent interests and evolving preferences, which are keys to achieve accurate and timely recommendations. In this paper, we focus on this special \textit{time-sensitive cold-start problem}, which is critical for a recommender system to maintain its user base. We propose to dynamically factorize a user's (latent) preference into time-specific and time-evolving representations in order to capture the time-specific and time-evolving factors from both the historical and current user-item interactions. For example, the Netflix dataset consists of a user's interactions with a large movie set in the form of user ratings. The variation of movie ratings from the same user may be affected by the change of user preference, rating criteria, or other (unknown) factors. 
In addition, a user's preference for different genres may evolve over multiple periods, as shown in Figure~\ref{time_evol}, which corresponds to the proportion of different factors in the time-evolving representation discovered by our proposed model.

The proposed model consists of two distinct modules: a \textit{meta-learning module} and a \textit{recurrent module}.  The former aims to capture time-specific latent factors through limited interaction data by leveraging the shared knowledge learned from other users. The latter aims to capture time-evolving latent factors by nesting a recurrent neural network, and it can be jointly optimized with the meta-learning module through the model-agnostic meta-learning approach \cite{finn2017model}.
Finally, we seamlessly integrate the two modules by merging the time-specific and time-evolving factors to form the user representation.
This user representation further interacts with an item embedding (which is also optimized during model training) to provide the final recommendations.  Our experimental results clearly show that the proposed model makes timely recommendations that closely resemble the dynamically changed user ratings as a result of effectively integrating the complementary factors capturing the user preferences.

The {\bf main contributions} of this paper are five-fold: (i) the {\bf first work} to formulate the time-sensitive cold-start problem that is critical to maintain the user base of a recommender system; (ii) a {\bf novel integrated recommendation framework} to model sparse dynamic user-item interactions and extract time-evolving and time-specific factors of user preferences simultaneously; (iii) a {\bf time-sensitive meta-learning module} to effectively handle user cold-start problems by leveraging knowledge shared across multiple users from the current recommendation period to adapt to any specific user's case using limited interaction information, (iv) a {\bf time-evolving recurrent module} to effectively capture the gradual shift of users' preferences over time, and (v) an {\bf integrated training process} that combines these two models to simultaneously learn time-specific and time-evolving factors and optimizes the item embedding.

We conduct extensive experiments over multiple real-world datasets and compare with representative state-of-art dynamic and meta-learning-based recommender systems to demonstrate the effectiveness of the proposed model.

\section{Related Work}

\paragraph{Matrix Factorization Models.} Matrix factorization is a commonly used collaborative filtering approach that characterizes users and items through latent factors inferred from item rating patterns \cite{koren2009matrix}. Using singular value decomposition (SVD) for recommendation is popularized by \textit{Simon Funk} \cite{funk_2006} in the Netflix prize competition and a probabilistic version is introduced by \cite{mnih2008probabilistic}. SVD is extended to SVD++ \cite{koren2008factorization} for processing implicit feedback. 

The above static models do not include the important temporal information, which should be considered for analyzing user-item interactions in a time-sensitive setting. Dynamic matrix factorization has been developed to address the issue, which allows latent features to change with time. Some works introduce time-specific factors, such as timeSVD++, which uses additive bias to model user-related temporal changes~\cite{timeSVD}. 
There have been works that employ Gaussian state-space models to introduce time-evolving factors with a one-way Kalman filter~\cite{sun2014collaborative,gultekin2014collaborative}. To handle implicit data, Sahoo et al. propose an extension to the hidden Markov model~\cite{sahoo2012hidden}, where clicks are drawn from a negative binomial distribution and Charlin et al. \cite{dpf} introduces a Gaussian state-space model with the Poisson emission. 
However, matrix-factorization-based algorithms may suffer from limited expressive power and may not capture the complex nature of user-item interactions. Instead, our model captures both dynamic and static user preferences.

\paragraph{Deep Learning Models.} Recent works in recommender systems \cite{cheng2016wide,guo2017deepfm,zhou2019deep} utilize deep learning to provide better recommendations. Cheng et al. \cite{cheng2016wide} propose to jointly train wide linear models and deep neural networks to combine the benefits of memorization and generalization.
Similarly, DeepFM \cite{guo2017deepfm} integrates the power of deep learning and factorization machines models to learn low- and high-order feature interactions simultaneously from the input. DIEN \cite{zhou2019deep} formulates interest evolution network as a deep learning model to capture latent temporal interests and evolving interests for better recommendations.
These models are very sensitive to the features and might need important features information and large datasets while training, which is not necessary for our model.

\paragraph{Graph-Based Models.} Another popular line of recommendation systems is graph-based models. A graph captures high-order user-item interactions through an iterative process to provide effective recommendations \cite{guo2020survey}. Users and items are represented as a bipartite graph in \cite{berg2017graph} and links are predicted to make recommendations. 
Similarly, a graph-based framework called Neural Graph Collaborative Filtering (NGCF) \cite{wang2019neural} explicitly encodes the collaborative signal in the form of high-order connectivities in a user-item bipartite graph via embedding propagation. However, these methods are unable to capture long-term user preferences or deal with cold-start problems. 

\paragraph{Sequential Models.} Sequential models understand the sequential user behaviors via user-item interactions, and model the evolution of users’ preferences and item popularity over time \cite{wang2019sequential,fang2020deep}. Tang et al. \cite{tang2018personalized} utilizes convolutional sequence embedding to capture union level and point level contributions of historical items via horizontal and vertical filters and provides top-N sequential recommendations. Similarly, Kang et al. \cite{kang2018self} introduce a self-attentive mechanism to handle both long and short-term user preferences in a sequential setting. 
Sequential models focus on users' evolving preferences (i.e., recent interactions) but largely neglect long-term users' preferences. In contrast, our model dynamically captures long-term time-evolving preference via RNN and time-specific users preferences through meta-learning to alleviate cold-start problems.

\paragraph{Meta-learning Models.} The user-item interaction data is usually sparse because a user may only interact with a few items within the large item pool. In such cases, making recommendations can be viewed as a few-shot learning problem. Meta-learning \cite{schmidhuber1987evolutionary,bengio1992optimization} is recently becoming a popular few-shot learning approach that learns from similar tasks and can generalize quickly and efficiently for the few-shot unseen new tasks. Finn et al. \cite{finn2017model} propose a model-agnostic meta-learning model that learns global parameters from a large number of tasks and performs as a good generalization on a new task that has few samples utilizing the few steps of gradients. To address the cold-start problem in item recommendation, Vartak et al. \cite{vartak2017meta} introduce a meta-learning strategy
that focuses on the items-cold-start problem to recommend cold-start items considering that items arrive continuously in the Twitter Timeline. 
Similarly, recent work based on meta-learning is done to estimate user preferences in \cite{lee2019melu} and scenario-specific recommendation in \cite{du2019sequential}. Both works use the information of users and items to generate user and item embeddings. Meta-learning is specifically utilized to learn heterogeneous information networks to alleviate cold-start problems \cite{lu2020meta}. Similarly, meta-learning approachs are utilized in graphs \cite{wei2020fast,xie2021long} to quickly adapt to new sub-graph that alleviate the cold-start problem. Also, meta-learning is applied in click-through rate (CTR) prediction \cite{pan2019warm} where desirable embeddings for the new ads are generated via meta-learner. These embedding methods are designed for a static setting, making them not applicable to learn user preferences evolving over time. In our work, we simultaneously learn time-evolving and user-specific preferences to provide accurate and timely recommendations for time-specific cold-start users with very limited recent interaction data.

 \section{The Dynamic Meta-Learning Recommendation Model}
\paragraph{Problem Settings.} We propose a dynamic recommendation model, where the input data is represented as $\{\mathcal{U}, \mathcal{I}, \mathcal{H}\}$, $\mathcal{U}$ is the user set, $\mathcal{I}$ is the item set, and $\mathcal{H}$ is the set of time periods.
A time period $t \in \mathcal{H}$ defines a particular time interval where the interactions for user $u$  with the item $i$ are aggregated based on timestamps. 
We perform recommendation in each time period with a recommendation function as
\begin{equation}
f_{\theta_u^t, \omega}^t(i) = \hat{r}^t_{(u,i)} \qquad \forall u \in \mathcal{U}, i \in \mathcal{I}, t \in \mathcal{H}
\end{equation}
where $\hat{r}^t_{(u,i)}$ is the recommended score for item $i$ assigned by user $u$ at period $t$, $\theta_u^t$ is user latent factor at time $t$, and $\omega$ is the parameter of the recurrent neural network module. The goal of a recommender system is to predict the recommendation scores that can accurately capture a user's true preference on items over time so that the recommended items are likely to be adopted by the users.

We formulate dynamic recommendations as a few-shot regression problem in the meta-learning setting. Users are {\em dynamically} partitioned into {\em meta-train} and {\em meta-test} sets based on their interactions in the current recommendation period $t$. 
In particular, the meta-train user set includes users with sufficient interactions, while the meta-test user set includes time-sensitive cold-start users who have only a few interactions in the current time period. Details for the train-test user splits are discussed in the experiment section. We consider a distribution over tasks $P(\mathcal{T})$, and each user is represented as a few-shot regression task $\mathcal{T}_u^t$ sampled from the given task distribution. In general, a task includes a $\textit{support set}\  \mathcal{S}_u^t $ and a $\textit{query set}\  \mathcal{Q}_u^t$. 
The support set includes a user's interactions at period $t$ where $k$ is interpreted as the number of shots (i.e., interactions). The query set includes the rest interactions of this user at period $t$.
\begin{equation}
\begin{aligned}
\mathcal{T}_u^t \sim P(\mathcal{T}): \quad
& \mathcal{S}_u^t = \{ (u,i_j),r^{t}_{(u,i_j)}\}_{j=1:k},  \\ 
&\mathcal{Q}_u^t = \{ (u,i_j),r^{t}_{(u,i_j)}\}_{j=k+1:N_t}
\end{aligned}
\end{equation}
where $N_t$ is the number of items a user interacted with at period $t$ and $r^{t}_{(u,i_j)}$ represents label (i.e. rating or count) from user $u$ to item $i_j$. We adopt episodic training \cite{vinyals2016matching}, where the training task mimics the test task for efficient meta-learning. The support set $\mathcal{S}$ in each episode works as the labeled training set on which the model is trained to minimize the loss over the query set $\mathcal{Q}$. The training process iterates episode by episode until convergence.

\begin{figure}
\centering
  \includegraphics[width=8cm,height=4.5cm]{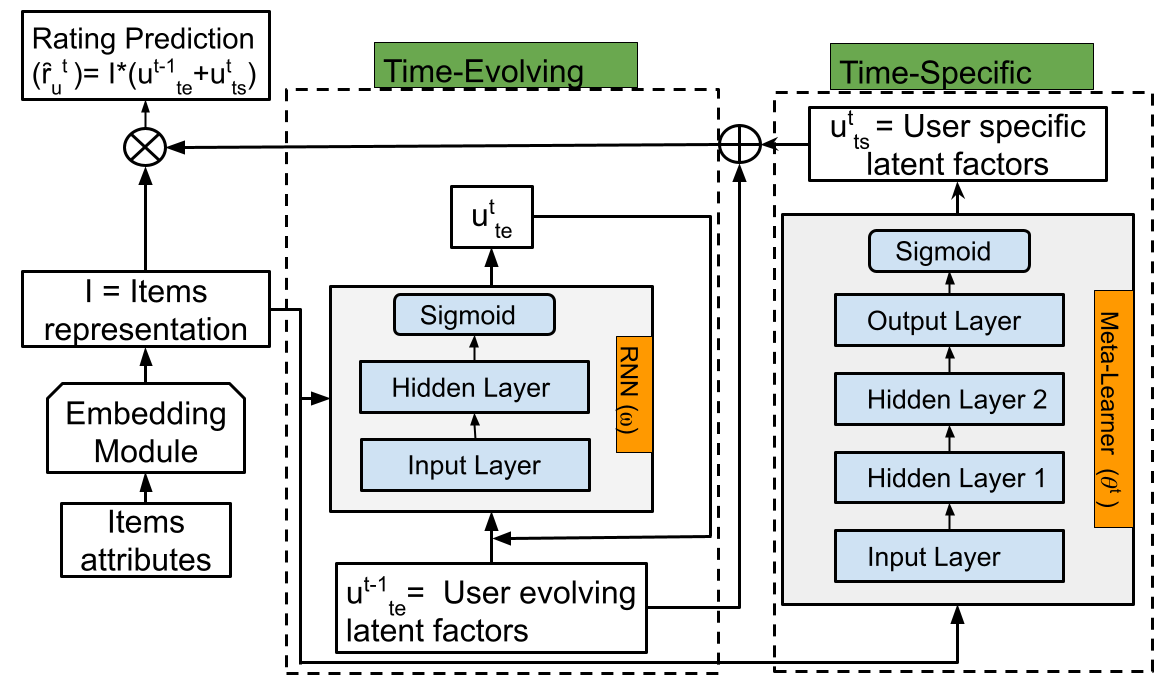}
  \caption{The proposed model captures time-evolving factors via a recurrent network module and time-specific factors through a meta-learning module. 
  \label{fig:arch}}
  \vspace{-4mm}
\end{figure}

\subsection{Model Overview}
To leverage item information such as text descriptions, our model generates an initial item representation ($I$) using an embedding matrix $E \in \mathbb{R}^{d\times m}$, where $m$ is the dimension of input item attributes, and $d$ is the dimension of embedding. The embedding is generated from item attributes following \cite{cheng2016wide,lee2019melu}. An item $i$ is first encoded as a binary vector $z_i \in \mathcal{R}^m$, where the corresponding index of item attributes is set to 1 and 0 otherwise. The binary vector is then transformed using the embedding matrix: $e_i = E z_i$. 
The embedding of all items can be stacked as: $I=[e_1,e_2,...,e_n]$ where $n$ is the total number of items. The embedding matrix $E$ will be optimized along with the model training process after the user latent factor is learned, and details are provided at the end of this section. 

Figure~\ref{fig:arch} shows that the proposed model consists of a time-specific meta-learning module and a time-evolving recurrent neural network module to generate time-specific user latent factors $u_{ts}^t$ and time-evolving latent factors $u_{te}^t$, both of which contribute to the final prediction $f^t(u,i)$.
Details of them are described in following sections. 
After both modules are trained, the model learns time-specific user factors $u_{ts}^t$ and time-evolving user factors $u_{te}^t$. These user factors are merged to interact with the item embedding to provide recommendations for the user. The recommendation for a user $u$ at the current period $t$ is denoted as a vector $\hat{r}_u^t$.

Making recommendations can be viewed as a regression problem. By using the mean square error (MSE) function, the loss for a specific user $u$ is formulated as:
\begin{equation}
\begin{aligned}
  & \mathcal{L}_{\mathcal{T}_u^t} [f^t_{\theta_u^t, \omega} ]   =\sum_{i} ||f^t_{\theta_u^t, \omega} (i)-r^t_{(u,i)} ||_2^2, \\
  & f^t_{\theta_u^t, \omega} (i)  = (u^t_{ts}+u^{t-1}_{te}) \cdot e_i 
\end{aligned}
\label{loss}
\end{equation}
where $r^t_{(u,i)}$ is user-item interaction (rating or count). The user representation is the vector sum of time-specific $u^t_{ts}$ and time-evolving user factors $u^{t-1}_{te}$ (as a compact single representation reducing the number of trainable parameters that helps to avoid overfitting), and the prediction is achieved by the dot product of $u$ and item embedding $i$.
Note that prediction for the current time includes time-specific user factors from current time period i.e. $u^t_{ts}$ and time-evolving user factors from the previous time period i.e. $u^{t-1}_{te}$. In general, dynamic recommender systems utilize the information from the previous period to predict for the next period, and we follow this standard setting.

The total loss is formed by aggregating all users in the meta-train set, regularized by the $L_2$ norm of key model parameters. Let $\theta_u^t$ and $\theta^t$ denote the local (i.e., user-specific) and global parameters of the time-specific meta-learning module, $\omega$ denote the parameters of the time-evolving recurrent neural network module. Training of a dynamic recommendation model can be formulated as: 
\begin{equation}
\begin{aligned}
\label{eq:regloss}
\arg\min_{\theta^t, \omega}   \quad &\sum_{_{\mathcal{T}_u^t \sim p(\mathcal{T})}} \mathcal{L}_{\mathcal{T}_u^t} [f^t_{\theta_u^t, \omega}] + \frac{\lambda}{2} (||\theta^t||^2_2 + ||\omega||^2_2),\\
&\theta_u^t  = \theta^t-\alpha \triangledown_{\theta^t} \mathcal{L}_{\mathcal{T}_u^t} [f^t_{\theta^t, \omega}]
\end{aligned}
\end{equation}
where $\theta_u^t$ is one gradient step from global parameter $\theta^t$ of the meta-learned time-specific module with $\alpha$ being the step size and $\lambda$ is the regularization parameter.

\subsection{Time-Specific Meta-Learning Module}
\label{metasection}
This module aims to capture time-specific user factors by only considering the information from the current recommendation period. The meta-learner takes input from the specific period, which is a different setting than the existing meta-learning-based recommender systems \cite{lee2019melu,du2019sequential}. In this way, the model can capture the latent factors associated with that specific period to provide more accurate and timely recommendations. 
We consider each user as a learning task. Our goal is to learn a meta parameter $\theta^{t}$ that represents a time-specific global user representation given the meta-training set. We follow the standard setting of few-shot learning \cite{finn2017model}, where the distribution over tasks is represented as $p(\mathcal{T})$. The model is trained iteratively by sampling tasks from $p(\mathcal{T})$. The meta-learning module generates time-specific user latent factors ($u^t_{ts}$) as:
\begin{equation}
u^t_{ts}=f_{meta}^t (\mathcal{T}_u^t;\theta^t)
\label{ts}
\end{equation}
where $\mathcal{T}_u^t$ represents task of a user $u$ at period $t$.
The task $\mathcal{T}_u^t$ includes $\mathcal{S}_u^t$  and $\mathcal{Q}_u^t$. We first pass $\mathcal{S}_u^t$ into $f_{meta}^t$ to adapt user-specific model parameter $\theta_u^t$ from the global user model parameter $\theta^t$ and then we provide $\mathcal{Q}_u^t$ into the $f_{meta}^t$ to generate time-specific user factors ($u^t_{ts}$).

We apply an optimization-based meta-learning approach \cite{finn2017model} to learn time-specific user factors, as shown in Figure~\ref{fig:arch}. 
The meta-learning network consists of one input layer, two fully connected hidden layers, and one output layer. The first and second hidden layers have 128 and 64 hidden units with ReLU activation, while the last layer estimates time-specific user factors with a linear function followed by sigmoid activation. The input to the meta-learning model is the item embedding for the users on a particular period.
Algorithm~\ref{alg:training} shows the training process that learns the model parameters. For the time-specific module, the local update (line 7) is done for the user specific parameter, which is achieved by one or more gradients from the global parameter:
\begin{align}
\theta_{u}^t=\theta^t -\alpha \nabla_{\theta^t}\mathcal{L}_{\mathcal{T}_u^t} [f^t_{\theta^t, \omega} ]
\label{localupdate}
\end{align}
In this update, the loss function is computed with the support set. Similarly, the global update (line 11) is conducted with the new item interactions of each user from the query set for the meta update:
\begin{align}
\theta^t=\theta^t -\beta \nabla_{\theta^t}  \sum_{\mathcal{T}_u^t \sim p(\mathcal{T})}  \mathcal{L}_{\mathcal{T}_u^t} [f^t_{\theta_u^t, \omega} ]
\label{globalupdate}
\end{align}
This process continues to find a good global parameter shared by all users in each period.

\subsection{Time-Evolving Module}
\label{rnnsection}
User preferences usually change dynamically over time. By capturing the time-evolving factors and integrating them with the time-specific factor, the proposed model can recover the user's true preference more accurately. To this end, we formulate time-evolving user factors ($u^t_{te}$) for each user using a nested recurrent neural network (RNN):
\begin{equation}
u^t_{te}=f^t_{rnn}(u_{te}^{t-1}, D_u^t;\omega)
\label{rnn}
\end{equation}
where ${D}_u^t$ is the set of items that user $u$ interacted with at time $t$, $u_{te}^{t-1}$ is the previous time period time-evolving user factors, and $\omega$ is the network parameter. Notice that the input and output of the RNN are both latent variables instead of observations. We use SGD to update the parameter of RNN:
\begin{equation}
\omega = \omega -\gamma \nabla_{\omega} ( \mathcal{L}_{\mathcal{T}_u^t}  [f^t_{\theta_u^t, \omega} ] + \frac{\lambda}{2} ||\omega||^2_2 )
\label{rnnupdate}
\end{equation}
where $\gamma$ is the step size. 
As shown in the time-evolving module of Figure 2, the vector representation of a hidden layer $u^{t}_{te}$ is a time-evolving factor of user $u$ at period $t$ and helps to propagate influence from the previous period to the next period \cite{zhang2014sequential}. The updates of time-specific user factors through meta-learning and time-evolving user factors through nested RNN are summarized in Algorithm~\ref{alg:training}. The recommendation process is summarized in Algorithm~\ref{alg:rec}.

\begin{algorithm}[htpb!]
\caption{\label{alg:training} Model training}
\begin{algorithmic}[1]
\Require Set of time periods: $\mathcal{H}$
\Require Hyperparameters:  $\alpha, \beta, \gamma$
\For {$t \in \mathcal{H}$}
\State Initialize meta learner, $\theta^t$
\While {not converge}
\State Sample tasks $\mathcal{T}_u^t \sim p(\mathcal{T})$
\For {all $\mathcal{T}_u^t$}
\State Sample support set $\mathcal{S}_u^{t}$ for local update
\State Perform local update with $\mathcal{S}_u^{t}$  for time-specific module using Equation~\eqref{localupdate}
\State Sample query set $\mathcal{Q}_u^{t}$ for meta update
\State Update time-evolving module with $\mathcal{D}_u^{t-1}$ using Equation~\eqref{rnnupdate}

\EndFor
\State Perform meta update with $\mathcal{Q}_u^{t} $ for time-specific module using Equation~\eqref{globalupdate}
\EndWhile
\EndFor
\end{algorithmic}
\end{algorithm}

\begin{algorithm}[t!]
\caption{Recommendation for time-specific cold-start users}
\label{alg:rec}
\begin{algorithmic}[1]
\Require Trained meta parameter $\theta^t$, RNN parameter $\omega$, recommendation time period $t$
\State Identify cold-start user set for $t$
\For {each user $u$ in the set}
\State  
Form support set $\mathcal{S}_u^{t}$ from current interactions
\State Perform local update with $\mathcal{S}_u^{t}$ for time-specific module using Equation~\eqref{localupdate}
\State Compute user factors using Equations~\eqref{ts} and ~\eqref{rnn}
\State Make recommendation using Equation~\eqref{loss}
\EndFor
\end{algorithmic}
\end{algorithm}

\paragraph{Joint Item Embedding Optimization.}
Let $\mathcal{L}_{emb}$ denote a differentiable loss function used to train the embedding matrix $E$.
And let $\mathcal{G}$ denote the decoding module, which is followed by attribute-wise sigmoid transformation:
\begin{equation}
\begin{aligned}
d_i &=\mathcal{G}(E z_i), \quad [\hat{z}_i]_j = \text{sigmoid}(\eta_j^T d_i) \\
&=  \frac{1}{1+ \exp{(-\eta_j^T d_i)} } \quad \forall j\in \{1,...,K\}
\end{aligned}
\end{equation}
where $z_i$ is the original item representation in a binary vector, $K$ is the length of $z_i$, $E$ is the embedding matrix, $d_i$ is the decoded item representation, $\eta$ is the parameter for attribute-wise Sigmoid transformation, and $\hat{z}_i$ is the recovered item representation. The loss function for learning the embedding matrix is a negative log-likelihood and is represented as:
\begin{equation}
\begin{aligned}
\mathcal{L}_{emb}&= -\sum_{i\in I} \sum_{j} [z_i]_j \log{[\hat{z}_i]_j}
\end{aligned}
\end{equation}
Other designs for item embedding with a differentiable loss function can also be applied.

To jointly train the embedding matrix, time-specific and time-evolving modules, we combine those losses, and optimize the total loss with respect to $\theta^t$, $\omega$ and $E$.
\begin{equation}
\begin{aligned}
&\arg\min_{\theta^t, \omega,E}  \sum_{_{\mathcal{T}_u^t \sim p(\mathcal{T})}} \mathcal{L}_{\mathcal{T}_u^t} [f^t_{\theta_u^t, \omega,E}] + \xi \mathcal{L}_{emb} +\\ &\frac{\lambda}{2} (||\theta^t||^2_2 + ||\omega||^2_2) \\
&\theta_u^t  = \theta^t-\alpha \triangledown_{\theta^t} \mathcal{L}_{\mathcal{T}_u^t} [f^t_{\theta^t, \omega, E}]
\end{aligned}
\end{equation}
where $\xi$ is the weight to be tuned. If $\xi$ is set to a very large value, matrix $E$ is determined only by $\mathcal{L}_{emb}$.

Note that the encoded item embedding is used for both modules. By fixing $\theta_u^t$ and $\omega$, it is possible to back-propagate and calculate the gradient with respect to $E$, which is updated in each task as
\begin{equation}
\begin{aligned}
E &=E -\gamma \nabla_{E} (\mathcal{L}_{\mathcal{T}_u^t}  [f^t_{\theta_u^t, \omega,E} ] + \mathcal{L}_{emb})
\end{aligned}
\label{eupdate}
\end{equation}

Also notice that $\mathcal{L}_{emb}$ is not dependent on  $\theta^t$ and $\omega$. Therefore, when embedding matrix is fixed, the loss function reduces to Eq~\eqref{eq:regloss}, and $\theta^t$ and $\omega$ are updated without considering $\mathcal{L}_{emb}$. 


\section{Experiments}
We conduct experiments on two movie datasets: {\em Netflix} and {\em MovieLens-1M}  that consist of users' ratings of movies as explicit feedback and one music dataset: {\em Last.fm} that consists of users' play counts of music tracks as implicit feedback. Besides reporting the overall recommendation performance and comparing with state-of-the-art baselines, we also investigate key properties of the model, including: (1) each module's performance when used in isolation, (2) impact of varying time period lengths, (3) recommendation performance with no interactions in the current period, and (4) impact of hyper-parameters in the model. Experimental settings, datasets details, and analysis of the above key properties are presented in the Appendix \cite{ourPaper}.


\paragraph{Methods for Comparison.} For comparison, we include matrix factorization based static and dynamic models, deep learning based models, graph models, sequential models, and meta-learning models:
\begin{itemize}[noitemsep,topsep=2pt]
\item \textit{Matrix factorization (MF):} The standard MF model SVD++ \cite{koren2008factorization} that also exploits both explicit and implicit feedback is used here as a static baseline.

\item \textit{Dynamic models:} We use timeSVD++ \cite{timeSVD}, collaborative Kalman Filter (CKF) \cite{gultekin2014collaborative}, and dynamic Poisson factorization (DPF) \cite{dpf} as the time-evolving models.

\item \textit{Deep learning models:} We use Wide and deep \cite{cheng2016wide}, DeepFM \cite{guo2017deepfm} as static, and DIEN \cite{zhou2019deep} as a dynamic models for deep learning-based recommendation. However, most of them are developed for click-through rate prediction in their original forms.

\item \textit{Graph-based model:} 
Most graph-based models are designed for static settings. For comparison, we use graph convolutional matrix completion (GC-MC) \cite{berg2017graph}, which models recommendation as link prediction in the graph, and neural graph collaborative filtering (NGCF) \cite{wang2019neural} that utilizes embedding propagation over user-item graphs.

\item \textit{Sequential model:} 
We use Sequential Recommendation via Convolutional Sequence Embedding (Caser) \cite{tang2018personalized}, which models recommendation as a unified and flexible structure to capture both preferences and sequential patterns, and transformer-based sequential recommendation model (SASRec) \cite{kang2018self} in our comparison.

\item \textit{Meta-learning models:} We follow the model-agnostic meta-learning model (MAML) to implement the meta-learning model similar to MeLU \cite{lee2019melu}. 
We also compared with the meta-learning model in \cite{vartak2017meta} that focuses on item cold-start problem (referred to as ML-ICS). The model is also designed for the classification setting, so we have to make adjustments to fit into our context.

\end{itemize}

\paragraph{Evaluation Metrics.} For evaluation, we analyze the experimental results in terms of both the deviation of predicted values from the ground truth and the errors of the ranking sequences. We use Root Mean Squared Error (RMSE) and Normalized Discounted Cumulative Gain (NDCG) averaged across all test users. RMSE is usually reported for explicit data, while NDCG is usually reported for implicit data:
\begin{equation}
\begin{aligned}
\text{RMSE} =&\sqrt{\sum_{r_{u,i}\in O}(\hat{r}_{u,i}-r_{u,i})^2/|O|}, \\
\text{NDCG}_u =&\sum_n\frac{{rel}^{pred}_n}{\log_2 (1+n)} /\sum_n\frac{{rel}^{ideal}_n}{\log_2 (1+n)}
\end{aligned}
\end{equation}
where $O$ is the observation set for the test set and ${rel}_n$ is the relevancy of $n^{th}$ item in the ranking sequence for user $u$, which is binary for implicit data or the rating for explicit data. To penalize the negative feedback, we linearly mapped the ratings to a range of [-1,1]. The NDCG is the fraction of Discounted Cumulative Gain (DCG) of recommendation result over the ideal DCG.

\begin{figure}[htpb]
\vspace{-3mm}
\centering
\begin{subfigure}{.49\textwidth}
  \centering
  \includegraphics[width=0.45\linewidth,height=35mm]{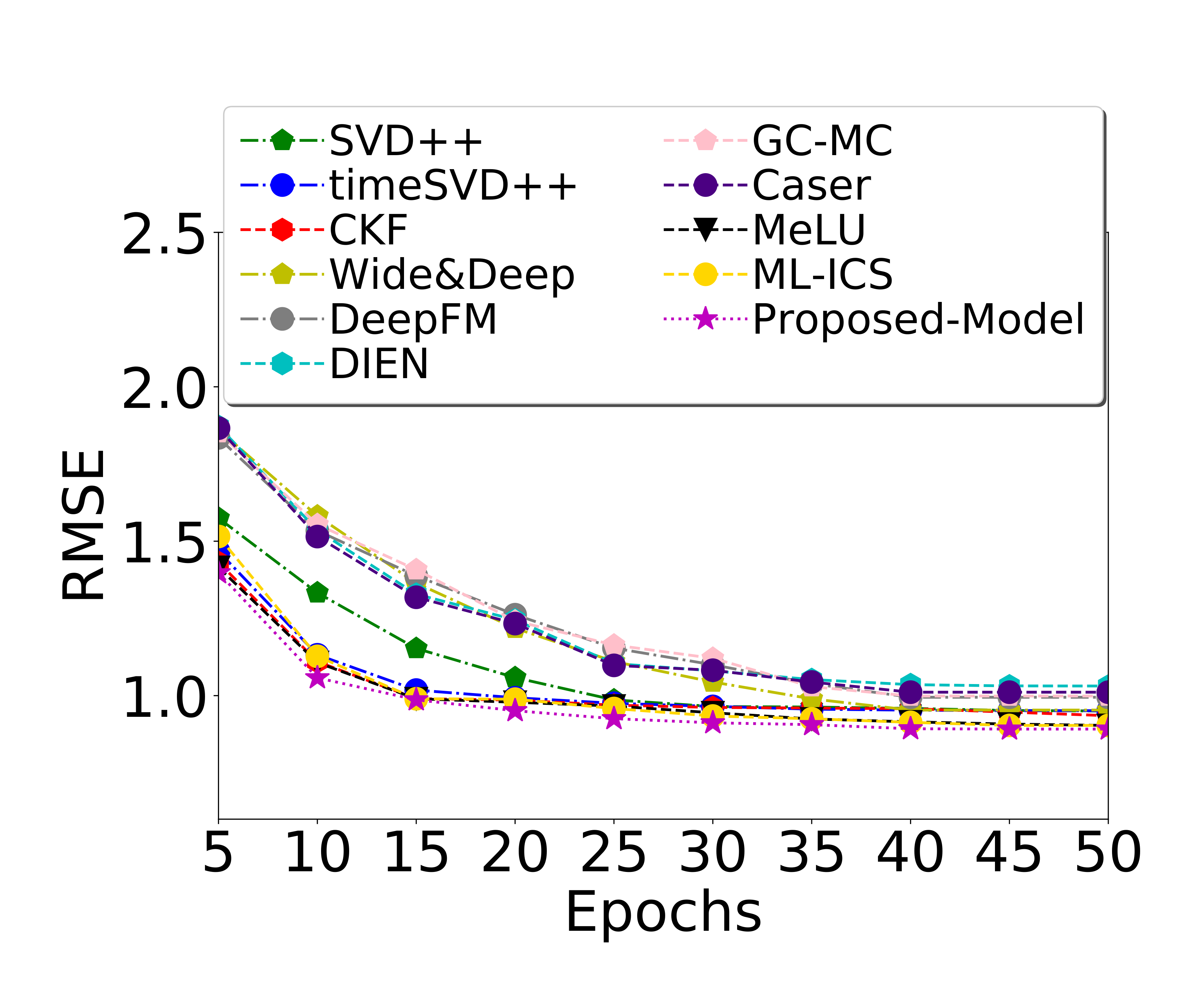}
  \includegraphics[width=0.45\linewidth,height=32mm]{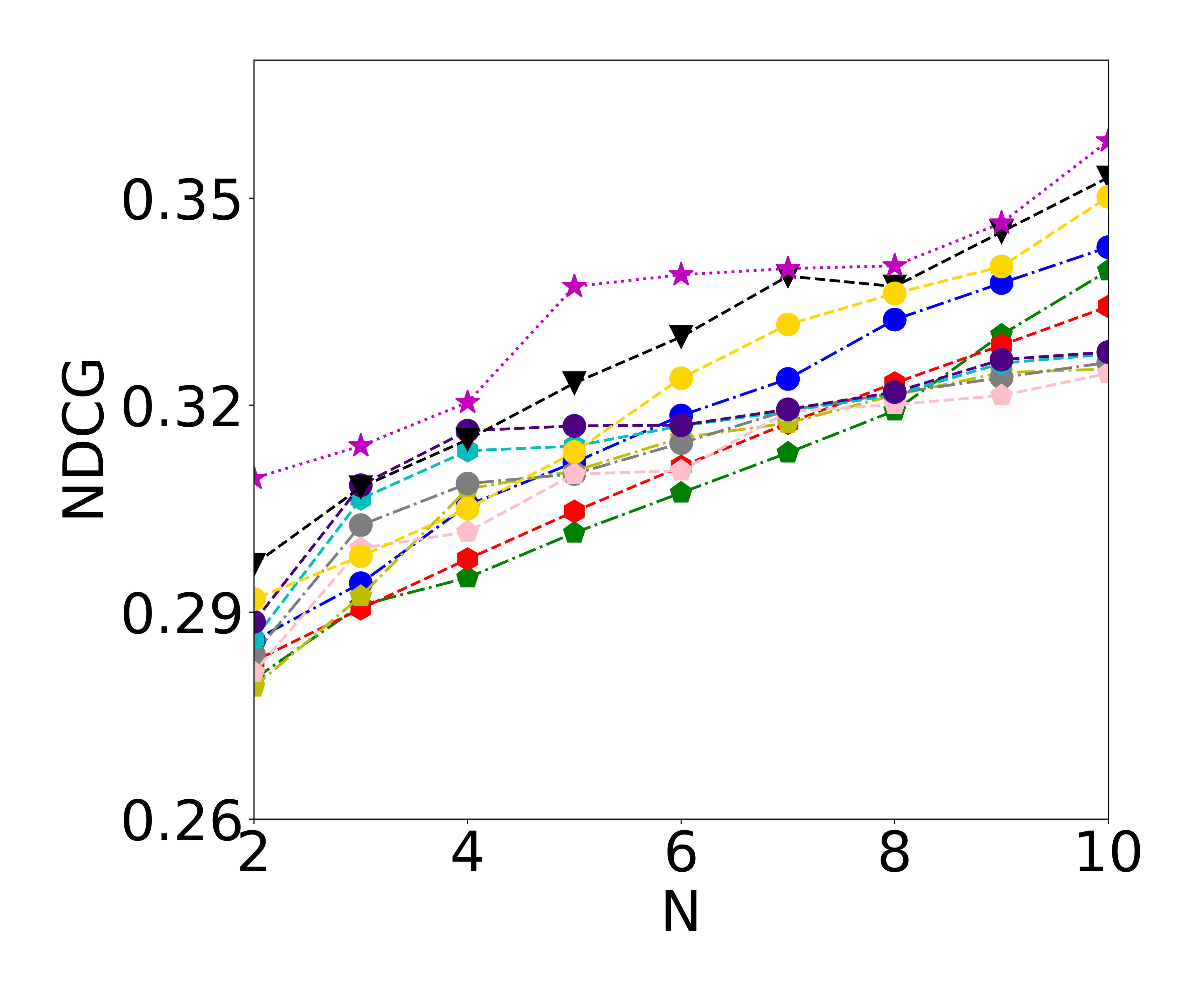}
  \vspace{-3mm}
    \caption{Netflix}
  \label{netflix}
\end{subfigure}
\vspace{-1mm}
\begin{subfigure}{.49\textwidth}
  \centering
  \includegraphics[width=0.45\linewidth, height=35mm]{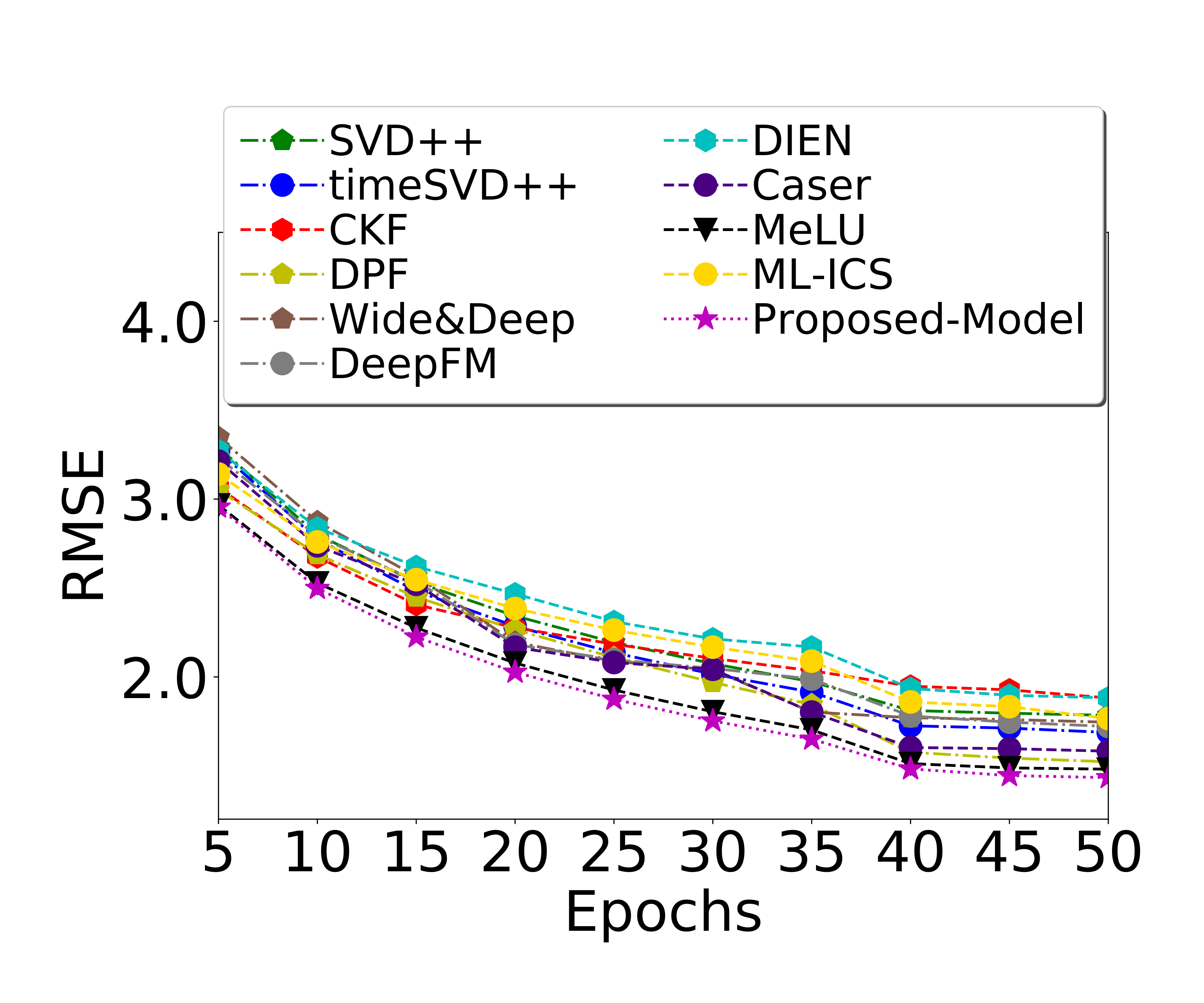}
  \includegraphics[width=0.45\linewidth, height=32mm]{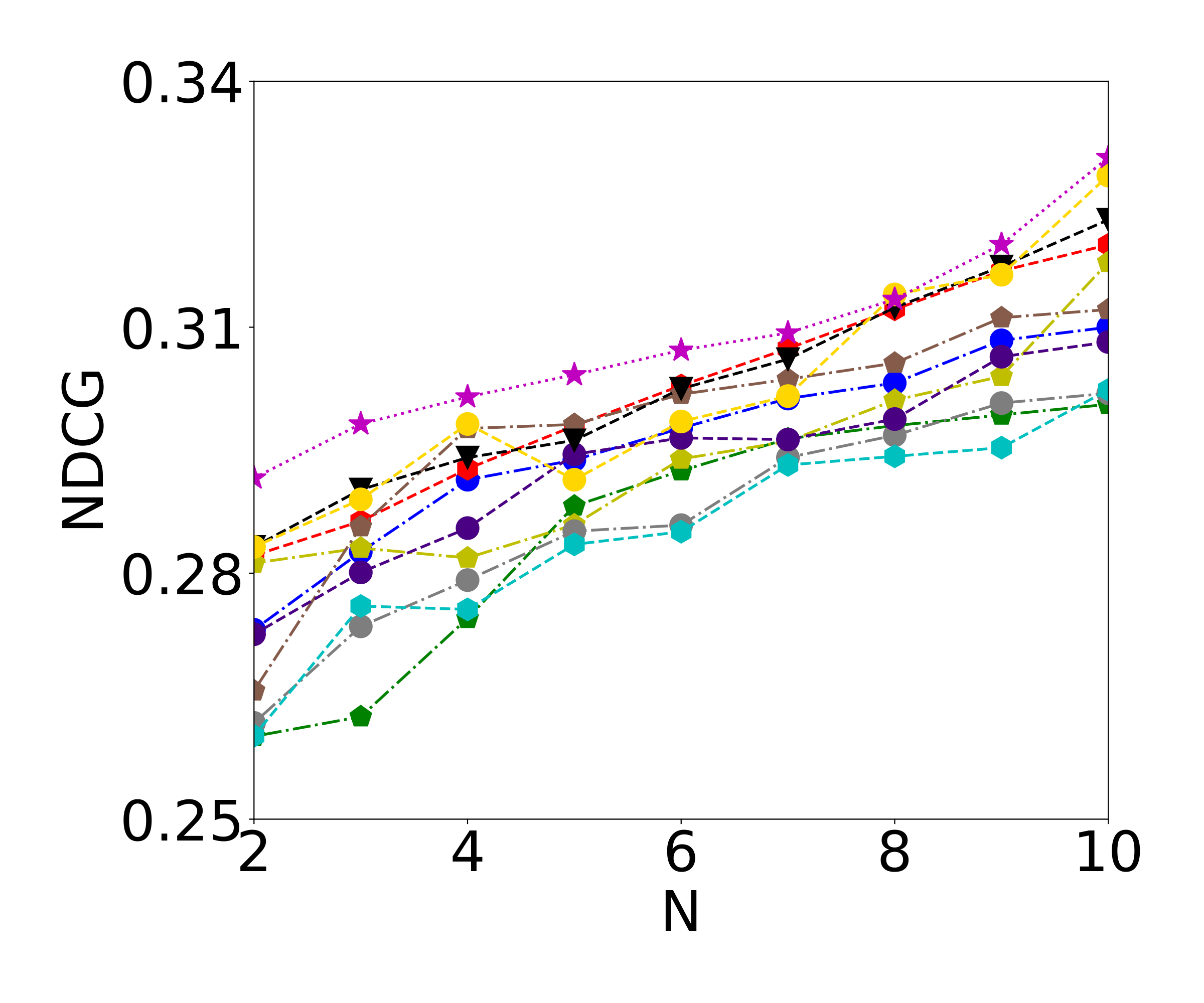}
  \vspace{-3mm}
  \caption{Last.fm}
  \label{last_fm}
\end{subfigure}
\vspace{-1mm}
\begin{subfigure}{.49\textwidth}
  \centering
  \includegraphics[width=0.45\linewidth,height=35mm]{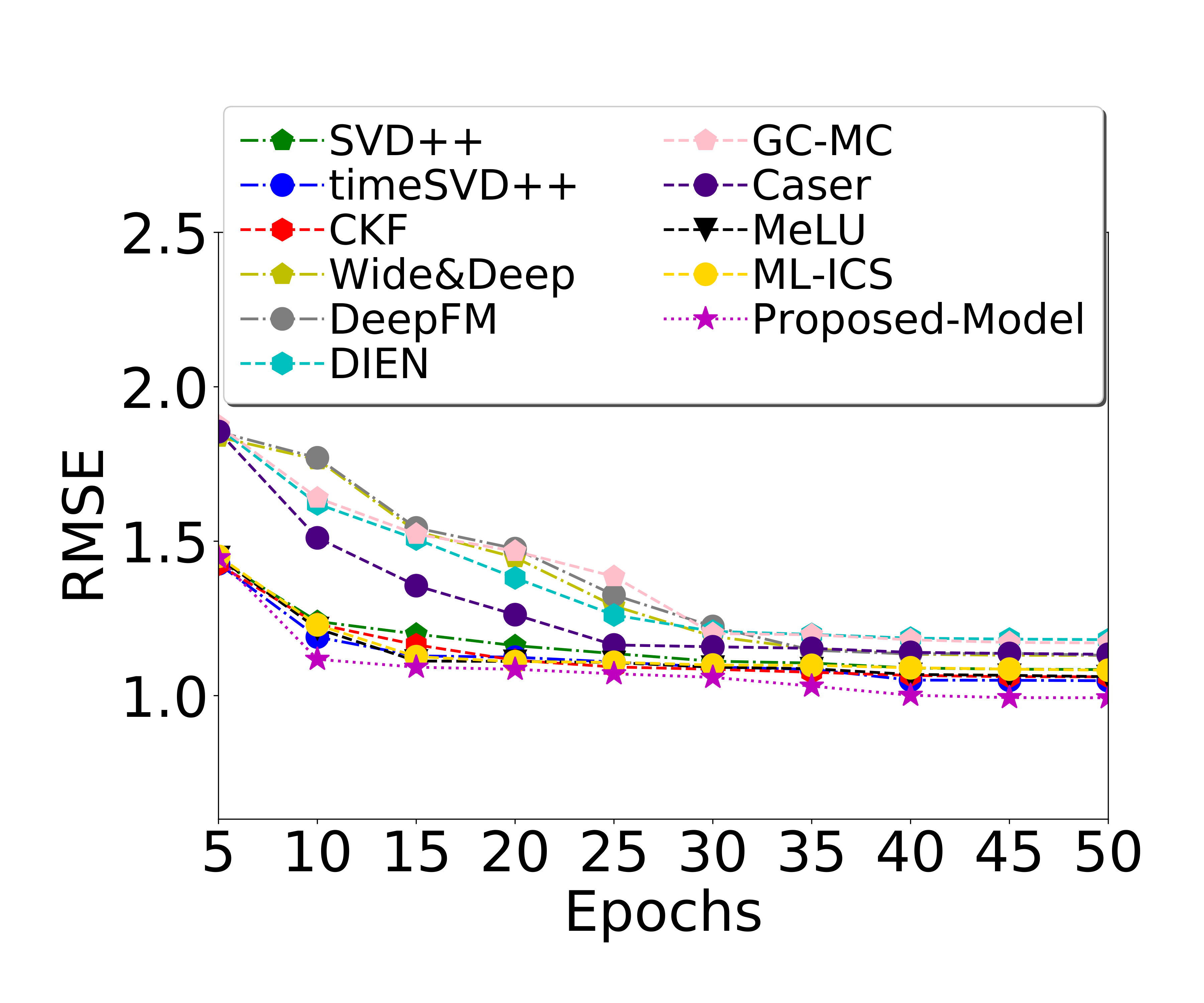}
  \includegraphics[width=0.45\linewidth,height=32mm]{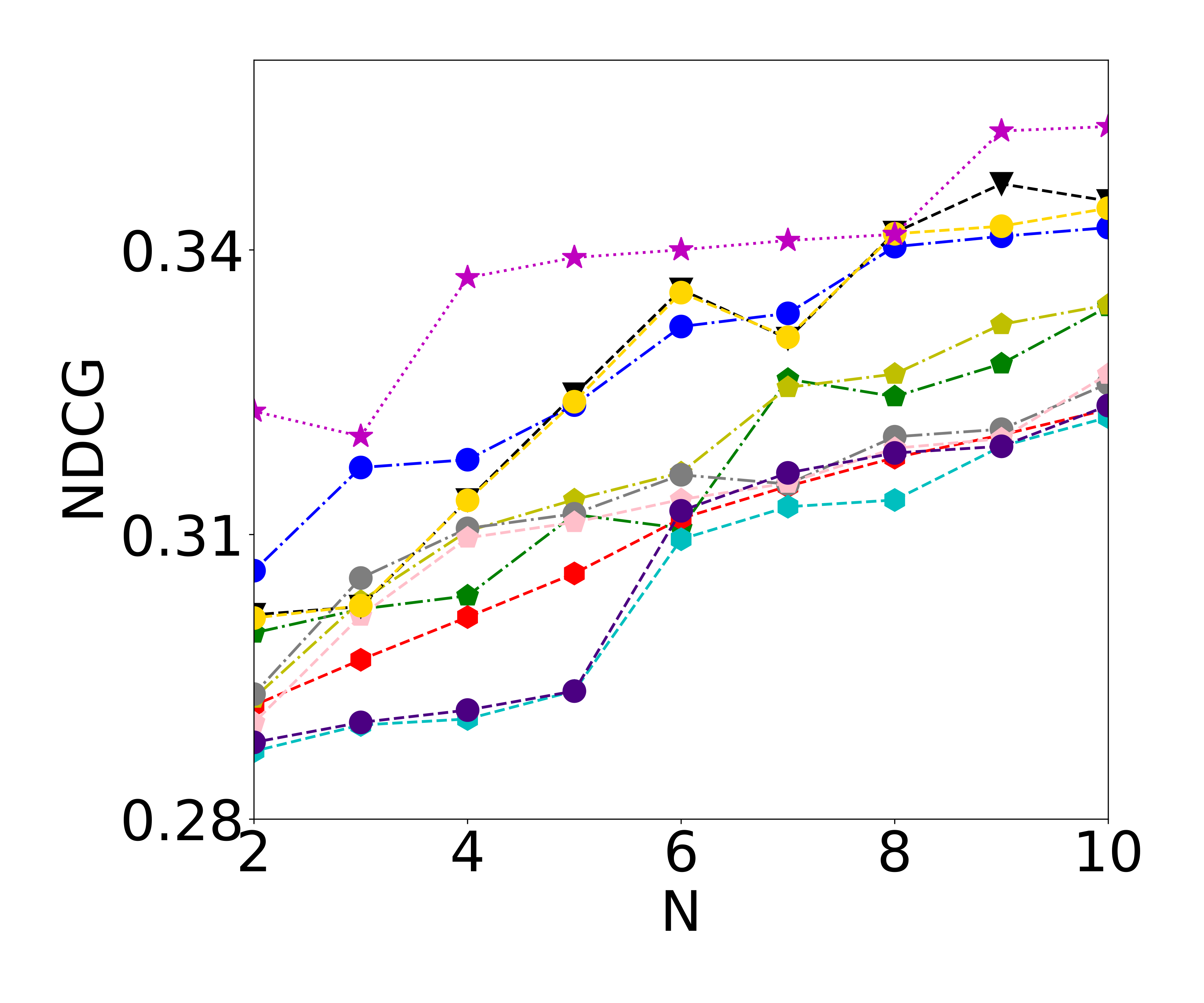}
  \vspace{-3mm}
    \caption{MovieLens-1M}
    \label{MovieLens-1M}
\end{subfigure}
\vspace{-1mm}
\caption{\label{fig:top}NDCG based on the top $N$ recommendations and RMSE based on the training epochs}
\vspace{-2mm}
\end{figure}

\begin{table*}[t]
\small
\begin{center}
 
\centering
 \vspace{-2mm}
\resizebox{.8\textwidth}{!}{ 
\begin{tabular}{|r|r|r|r|r|r|r|r|}
\hline
\multicolumn{1}{|c|}{\bf Category} & 
\multicolumn{1}{|c|}{\bf Model} & \multicolumn{2}{|c|}{\bf Netflix} & \multicolumn{2}{|c|}{\bf Last.fm} &
\multicolumn{2}{|c|}{\bf MovieLens-1M} \\
\cline{3-8}
\multicolumn{1}{|c|}{} &
\multicolumn{1}{|c|}{} &
 \multicolumn{1}{|c|}{\bf RMSE} & \multicolumn{1}{|c|}{\bf NDCG}  & \multicolumn{1}{|c|}{\bf RMSE} & \multicolumn{1}{|c|}{\bf NDCG}  & \multicolumn{1}{|c|}{\bf RMSE} & 
 \multicolumn{1}{|c|}{\bf NDCG}  \\
\hline 
MF&SVD++ &0.9797$\pm$0.03 &0.2915&1.7829$\pm$0.08 &0.2882  &1.0825$\pm$0.04 &0.3023 \\
\cline{1-8}
&timeSVD++ &0.9538$\pm$0.06 & 0.3115 &1.6912$\pm$0.11 & 0.2962&1.0483$\pm$0.03 & 0.3224  \\
Dynamic&CKF & 0.9337$\pm$0.04 &0.3130 &1.5316$\pm$0.32 &0.3018&1.0652$\pm$0.04 &0.3151 \\
&DPF & N/A & N/A &1.5227$\pm$0.43 & 0.3085&N/A & N/A  \\
\cline{1-8}
&Wide and Deep & 0.9904$\pm$0.04&0.2864 & 1.7253$\pm$0.22 & 0.2727 & 1.1364$\pm$0.06& 0.2932  \\
Deep Learning&DeepFM & 0.9811$\pm$0.03 & 0.2930 & 1.6815$\pm$0.21& 0.2971 &1.1723$\pm$0.05& 0.2882  \\
&DIEN & 1.0345$\pm$0.04 & 0.2832 &  1.9225$\pm$0.26&0.2714 & 1.1872$\pm$0.14& 0.2843\\
\cline{1-8}
Graph&GC-MC &1.0760$\pm$0.03 &0.2901&N/A &N/A &1.1704$\pm$0.08 &0.2913 \\
&NGCF&1.0321$\pm$0.03 &0.3026&1.5612$\pm$0.23 &0.2896 &1.1216$\pm$0.05 &0.3103\\
\cline{1-8}
Sequential&Caser&1.0124$\pm$0.03 &0.3101 &1.5824$\pm$0.31 &0.2931 &1.1339$\pm$0.08 & 0.3012\\
&SASRec&N/A &0.3246 &N/A &0.3103 &N/A & 0.3238\\

\cline{1-8}
Meta-Learning&MeLU & 0.9213$\pm$0.05 & 0.3232 & 1.2580$\pm$0.28 & 0.3122&1.0685$\pm$0.08 & 0.3214  \\
&ML-ICS & 0.9332$\pm$0.04 & 0.3173 & 1.2408$\pm$0.24 & 0.3142&1.0845$\pm$0.06 & 0.3244\\
\cline{1-8}
Proposed& \bf{Ours} & {\bf 0.8925$\pm$0.03} & {\bf 0.3472} & {\bf 1.2203$\pm$0.16} & {\bf 0.3385}& {\bf 0.9945$\pm$0.08} & {\bf 0.3351}  \\

\hline
\end{tabular}}
\vspace{-3mm}
\caption{Recommendation Results (RMSE and NDCG)}
\label{tbl:rmse_ndcg}
\end{center}

\vspace{-4mm}

\end{table*}

\subsection{Recommendation Performance} 
The experimental results are shown in Figure~\ref{fig:top}. We evaluate NDCG based on the top $N$ recommendation list and RMSE based on the training epochs. The RMSE is stable after 30-40 epochs in all datasets.

The average results of NDCG and RMSE considering all periods with the range of deviation are shown in Table~\ref{tbl:rmse_ndcg}, for the Netflix, Last.fm, and MovieLens datasets, respectively. The proposed model clearly demonstrates the advantage of combining time-specific and time-evolving user factors that lead to a superior recommendation accuracy as compared with other competitive models. Both explicit and implicit datasets are highly sparse, and MF models' performance is poor due to the sparse interactions. Also, MF models suffer from cold-start problems, and thus their performances are fairly limited, as shown in Table~\ref{tbl:rmse_ndcg}. Similarly, deep learning models require sufficient training data and hence largely suffer in the few-shot recommendation setting. Moreover, these models might need extra side information, like user profile and item details, for better recommendations. For example, DIEN  needs cleverly chosen interest features like user behavior, and the absence of those features limits its performance, as shown in Table~\ref{tbl:rmse_ndcg}. Similarly, the poor performance of graph-based models in both movie datasets implies that these methods are insufficient to handle cold-start problems. 
Like other existing models, the performance of a sequential model is less effective for the cold-start users in all three datasets. The reason could be that the model is less effective in capturing long-term user preferences.
In contrast, the meta-learning approaches show better results by leveraging shared knowledge across the users. However, in the time-specific cold-start setting, test users have very limited interactions. In particular, the meta-learning model doesn't benefit from time-evolving aspects of the user interests, and thus underperforms the proposed model.

\begin{figure}[t!]
\centering
\begin{subfigure}{.25\textwidth}
  \centering
  \includegraphics[width=0.95\linewidth]{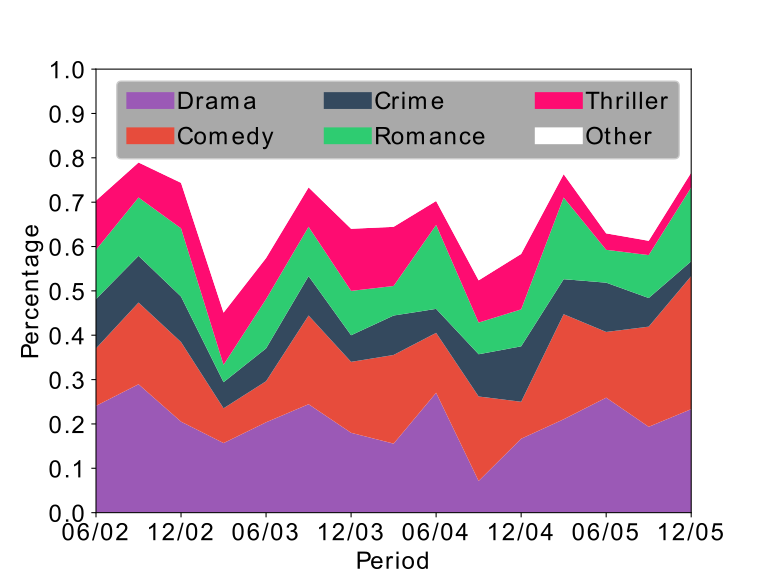}
  \caption{Ground Truth}

\end{subfigure}%
\begin{subfigure}{0.25\textwidth}
  \centering
  \includegraphics[width=0.95\linewidth]{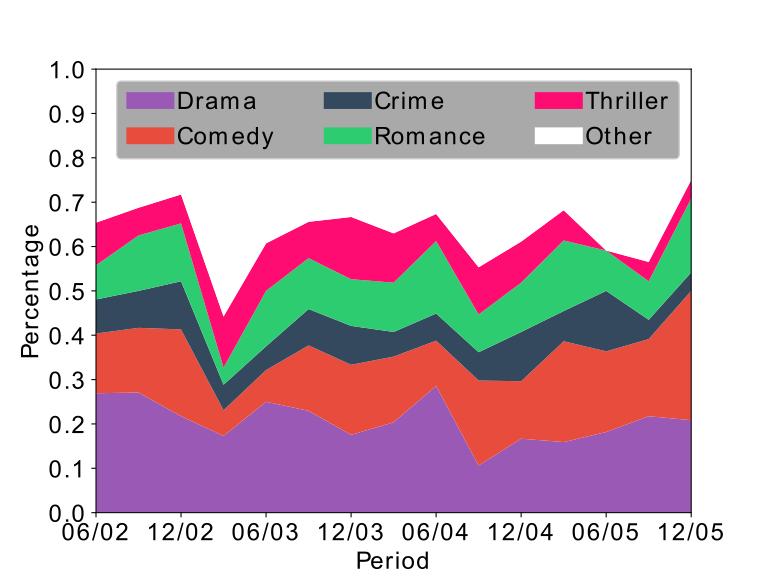}
  \caption{Recommended}

\end{subfigure}
\vspace{-3mm}
\caption{Dynamic trend of movie genres in Netflix: ground truth (a) vs model recommended (b) }
\label{dynamic_trend}
\vspace{-0mm}
\end{figure}

We use an illustrative example to further demonstrate how the proposed model effectively captures the underlying user interest and its evolution in Figure~\ref{dynamic_trend}. The recommended movie genres are compared to the user's favorite genres based on the provided true ratings. The result shows that the recommendation matches user's changing taste over time well. It is also interesting to see that the proposed model accurately detects some dramatic changes in user's ratings (e.g., from 12/02 to 06/03 and from 06/04 to 12/04), which were likely to be caused by some time-specific factors. 

We further present an example to show how the meta-learning module effectively captures time-specific factors in the form of popular trends in a specific period from the global user space and transfers the (meta) knowledge to the cold-start users with very limited interactions. Table~\ref{popular_items} demonstrates top-5 time-specific (period 4) popular movies learned by the meta-learning module, which shares that knowledge with the test users (i.e., two users are shown in Table~\ref{popular_items} and some time-specific popular movies like 'Best in Show' are recommended to them). 
This example demonstrates how our model provides effective recommendations to users by capturing time-specific factors.

\begin{table}[t]
\small
\centering
    \begin{tabular}{|p{2.5cm}|p{5.cm}|}
    \hline
    \textbf{Users}&\textbf{Movies}\\
    \hline
      Training users & ['Best in Show', 'Chicken Run', 'Sommersby', 'Bedrooms and Hallways', 'The Mod Squad'] \\
      \hline
      Test user (ID:5636)  & ['Mr. Mom', 'Best in Show', 'Shower', "We're No Angels", 'Groundhog Day']\\
      Test user (ID:5539)  &  ['Best in Show', 'An Ideal Husband', 'Life Is Beautiful', 'Breaking Away', 'Kramer vs. Kramer']\\
      \hline
\end{tabular}
\vspace{-3mm}
\caption{Time-specific popular movies learned and predicted by the meta-learning module}
\label{popular_items}
\vspace{-2mm}
\end{table}

\vspace{-2mm}\section{Conclusion}
In this paper, we formulate a novel time-sensitive cold-start problem and present a dynamic recommendation framework to address its unique challenges. The framework integrates a time-sensitive meta-learning module with a time-evolving recurrent module. The former handles the user cold-start problem by learning global knowledge among users from their interaction information in the current recommendation period. This module is jointly optimized with the time-evolving recurrent module that captures a user's gradually shifted preferences. A merged user representation is generated using the two modules' outputs and interacts with the item embedding to provide the final recommendations. 

\newpage
\section{Acknowledgements}
This research was supported in part by an NSF IIS award IIS-1814450, an ONR award N00014-18-1-2875, and by the Army Research Office under grant number W911NF-21-1-0236.
The views and conclusions contained in this paper are those of the authors and should not be interpreted as representing any funding agency.

\section{Ethical Impact}
The proposed model is generally applicable to many real-world applications where maintaining existing users' participation is critical, such as product recommendation in e-business and content recommendation in social media. Besides, this model can operate on multiple platforms such as health, education, and e-commerce to handle those sensitive users in the system. 
Since the proposed model does not rely on using users' profile information, it has the potential to improve the privacy protection of recommender systems.

\bibliography{main_bib}

\afterpage{\blankpage}
\newpage
\appendix

\begin{center}
 {\large \bf Appendix}    
\end{center}

\section{Summary of Notations}
\label{notation_section}
We summarize the major notations used throughout the paper in Table~\ref{tab:symbols}.
\begin{table} [htpb]
\caption{\label{tab:symbols} Summary of Notations}
\centering
\small
\vspace{-2mm}
\begin{tabular}{p{1.8cm}|p{5.2cm}}
\hline
\textbf{Notation} & \textbf{Description} \\
\hline
$u$, $i$ &  user and item \\\hline
$z_i$, $e_i$ & item $i$'s original encoding and embedding\\\hline
$\hat{r}^t_{(u,i)}$, $r^t_{(u,i)}$ & predicted and ground truth scores for user $u$ on item $i$ at period $t$ \\\hline
$u_{ts}^t$, $u_{te}^t$ & time-specific and time-evolving factors of user $u$ at period $t$ \\\hline
$\theta^t$, $\theta_u^t$ & meta and user-specific parameters of time-specific module at period $t$ \\\hline
$\omega$ & parameter of time-evolving module \\\hline
$\mathcal{S}_u^t$, $\mathcal{Q}_u^t$ & support and query sets in task corresponding to user $u$ at period $t$ \\\hline
${D}_u^t$ & items interacted with user $u$ at period $t$ \\
\hline
\end{tabular}
\end{table}

\section{Experimental Settings} 
We initialize both meta-learning and recurrent models with random initial values. Model learning rates are set through a grid search, and the Adam optimizer is applied with L2-regularization. In a dynamic meta-learning setting, the recurrent module takes historical interactions up to $t-1$ time period as an input while predicting for the next time period $t$, but the meta-learning module takes few recent interactions from the current time as a support set (e.g., in our setting, we have three months of period, so it takes $K$-shot interactions from the first month of the $t$ time period and remaining interactions of two months as query set). We split users into meta-train and meta-test sets. To make the problem more challenging, we consider time-sensitive users with few interactions in the current period as test users. We set the few-shot ($K=5$) to select the limited interactions for the support set.

For non-meta learning methods, the meta-train and meta-test split does not apply. To make a fair comparison, we first collect user interactions from period 1 to $t-1$ and then take the first few interactions ($K$) from the current time period $t$ to create the training set. The remaining interactions from time period $t$ are used for the test set. This mimics the few-shot problem for the current time period, and then a few interactions are available for the model to learn the user factors.

Different from explicit ratings that are constrained in a range, some implicit counts could take very large values. As a result, most of the models are not specifically designed for implicit data. To address this issue, we take a logarithm transformation on the Last.fm dataset to make the observed entries more balanced. Since DPF is specifically designed for counts data (only compared on Last.fm), we do not take logarithm transformation during training, but take such transformation on prediction and observation when calculating RMSE to make a fair comparison.

\section{ Datasets} The Netflix dataset has around 100 million interactions, 480,000 users, and nearly 18,000 movies rated between 1998 to 2005.
We preprocessed the dataset similar to \cite{li2011cross}, which consists of user-item interactions from 01/2002 to 12/2005. 
Movie attributes are taken from the IMDB website, in which we considered genres, directors, actors, movie descriptions, and overall movie ratings as the key movie attributes. The MovieLens-1M dataset includes 1M explicit feedback (i.e., ratings) made by 6,040 anonymous users on 3,900 distinct movies from 04/2000 to 02/2003. This dataset has very dense ratings in a particular time range. Thus we split datasets in a period of 6-months and got a total of 6 periods in contrast to the other two datasets, which have 16 periods considering a period of 3 months. We use the same preprocessing for this dataset as we did for the Netflix dataset. The Last.fm dataset is created by crawling the user interactions and track details from the Last.fm database. 
This dataset includes 12,902 unique tracks and 548 users from 01/2012 to 12/2015. Tracks are described with artists, tags, and summary information.

\section{Performance of Individual Modules}
The proposed model integrates two modules to capture time-specific and time-evolving latent factors. We study their contribution in detail to show how the user's time-specific interest and time-evolving preferences affect the overall recommendation performance.
\begin{itemize}
    \item {\bf Time-specific Meta-Learning (TS-ML) Module.} This module is specifically designed to capture users' time-specific interest in each period. Different from \cite{lee2019melu}, it only relies on time-specific data so that the proposed model is more generally applicable. 
  \item  {\bf Time-evolving RNN (TE-RNN) Module.} This module is designed to capture users' gradual shift of interest by utilizing historical interactions. 
\end{itemize}

\begin{table}[h]
\small
 \caption{Comparison of recommendation performance (Average RMSE and NDCG) using each module alone and the proposed integrated model for all three datasets }
 \vspace{-2mm}
 \centering
\begin{tabular}{|r|r|r|r|}
\hline
\multicolumn{1}{|c|}{\bf Dataset} & 
\multicolumn{1}{|c|}{\bf Model} & \multicolumn{1}{|c|}{\bf RMSE}& \multicolumn{1}{|c|}{\bf NDCG} \\
\hline 
&TS-ML &0.9380$\pm$0.02&0.3103 \\
Netflix&TE-RNN &0.9478$\pm$0.03 &0.3011\\
&Proposed &{\bf 0.8925$\pm$0.03}& {\bf0.3472} \\
\cline{1-4}
&TS-ML&1.2791$\pm$0.12 &0.3073\\
Last.fm&TE-RNN &1.9938$\pm$0.34&0.2910 \\
&Proposed &{\bf 1.2203$\pm$0.16}&{\bf0.3385} \\
\cline{1-4}
&TS-ML &1.0935$\pm$0.07&0.3144 \\
MovieLens-1M&TE-RNN &1.2505$\pm$0.13&0.3162 \\
&Proposed &{\bf 0.9945$\pm$0.08}&{\bf 0.3351} \\
\hline
\end{tabular}
\label{each_module}
\end{table}

Table~\ref{each_module} reports the recommendation performances of each module  and compares them with the proposed integrated model. First, each module performs reasonably well, and the recommendation results are comparable with some of the state-of-the-art baselines. It is also interesting to see that the meta-learning model outperforms the time-evolving module in all cases. This demonstrates the stronger impact of the time-specific factors when making recommendations to users. It also justifies the proposed meta-learning module's effectiveness that successfully captures these latent factors by learning the global trend from other users during a specific period. Finally, the proposed model that integrates both modules achieves the best recommendation performance, because it can capture both the time-evolving and time-specific factors.

\begin{figure}[h]
    \centering
    \includegraphics[height=3.5cm]{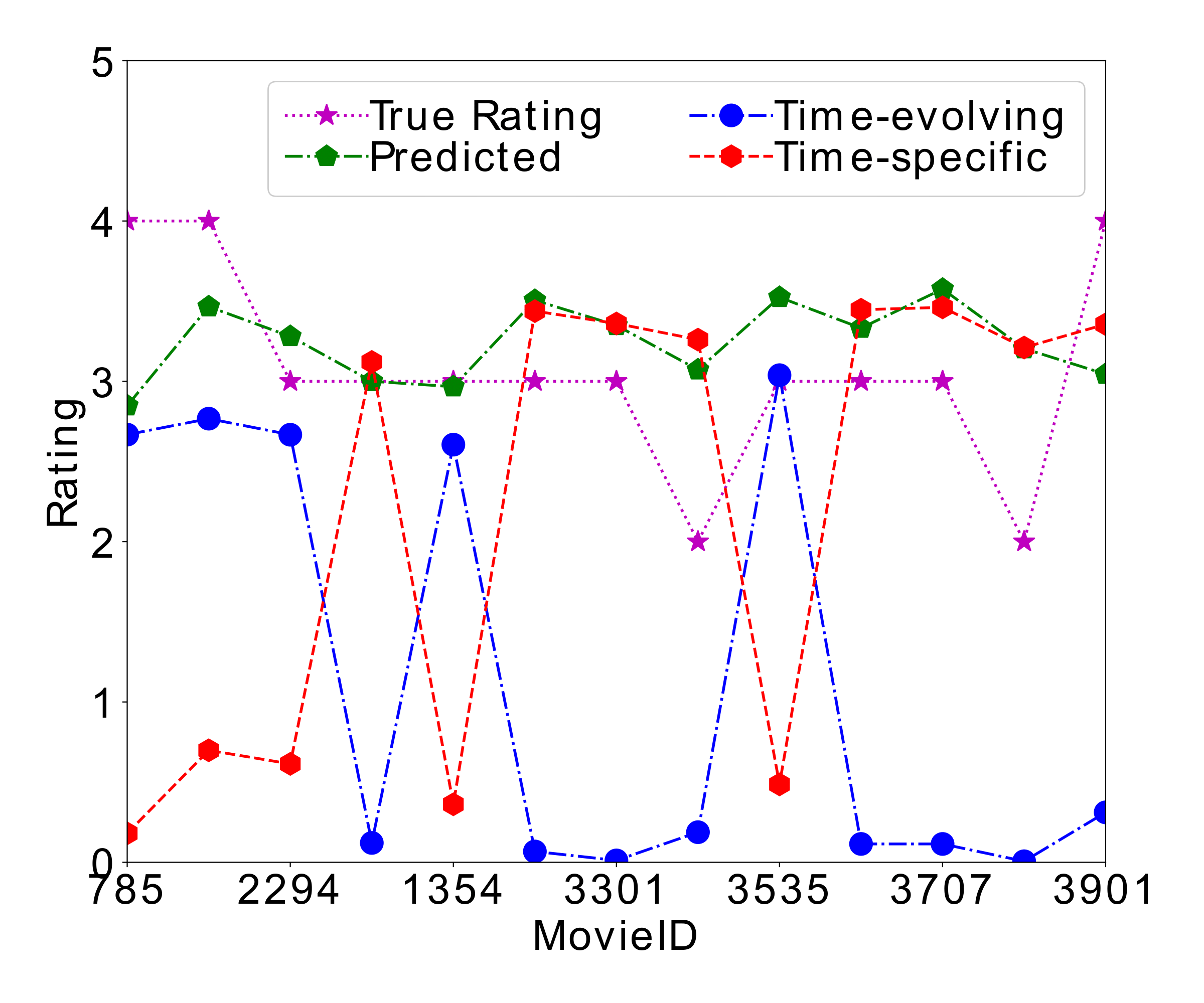}
    \vspace{-4mm}
    \caption{Contribution of time-evolving and time-specific modules to the recommendation of different movies, where they jointly contribute to the predicted ratings.}
    \label{combine_contribution}
    \vspace{-4mm}
\end{figure}

We further provide an illustrative example from the MovieLens-1M dataset for a user with ID:3462 to show each module's contribution to the final prediction. Figure~\ref{combine_contribution} shows that final predicted ratings are the combination of both modules, and they effectively complement each other to provide better final performance.

\section{Varied Recommendation Period Lengths}

In this section, we show a more fine-grained temporal analysis by varying the period length. We use three different period lengths: 1 month, 3 months, and 6 months, in the Netflix dataset and report the model performance. We also select one representative from each category of baseline models and report the RMSE and NDCG in Table~\ref{period_length}. The user base and the number of interactions per user are limited when we set the length to 1 month. Due to this, matrix factorization methods and deep learning methods suffer largely and have poor performance compared to the proposed model. In the case of 6 months, due to more interactions, deep learning models are performing better than other smaller period lengths. Also, we perform $K=5$-shots for meta-learning, and due to large interactions present in a 6-month period length, it is quite possible not to get those shots from the most recent interactions. This could hurt the performance of meta-learning, which achieves slightly lower performance than the 3-month period. Overall, in all period lengths, our model outperforms others with a noticeable margin.

\begin{table*}[t]
\small
 \caption{RMSE and NDCG in varying period lengths in months}
 \vspace{-2mm}
\centering
    \begin{tabular}{|r|r|r|r|r|r|r|r|}
\hline
\multicolumn{1}{|c|}{\bf Model} & \multicolumn{2}{|c|}{\bf 1} 
& \multicolumn{2}{|c|}{\bf 3} 
& \multicolumn{2}{|c|}{\bf 6} 
\\
\cline{2-7}
\multicolumn{1}{|c|}{} &
 \multicolumn{1}{|c|}{\bf RMSE} & \multicolumn{1}{|c|}{\bf NDCG}   &
 \multicolumn{1}{|c|}{\bf RMSE} & \multicolumn{1}{|c|}{\bf NDCG}  &
 \multicolumn{1}{|c|}{\bf RMSE} & \multicolumn{1}{|c|}{\bf NDCG} \\
\hline 
SVD++ &1.0923$\pm$0.04&0.2892 &0.9797$\pm$0.03 &0.2915& 0.9730$\pm$0.05&0.2901\\

timeSVD++ & 1.0742$\pm$0.04&0.3086 &0.9538$\pm$0.06 &0.3125&0.9641$\pm$0.04&0.31142\\
deepFM &1.1260$\pm$0.09&0.2817 &0.9811$\pm$0.04 &0.2930 &0.9585$\pm$0.06 &0.2891 \\
MeLU &1.0037$\pm$0.05&0.3138 &0.9213$\pm$0.05&0.3232&0.9414$\pm$0.04&0.3204\\
\textbf{Proposed} &\textbf{0.9864$\pm$0.06}&\textbf{0.32070} &\textbf{0.8925$\pm$0.03} &\textbf{0.3472} &\textbf{0.9226$\pm$0.03} 
 &\textbf{0.3317}\\
\hline
\end{tabular}
\label{period_length}
\vspace{-2mm}
\end{table*}

\section{No Interactions in the Current Period}

We further study the model performance where a user doesn't have any interaction in the current period. In this case, our time-specific module utilizes global user factors instead of user-specific factors due to lack of interaction, and hence no adaptation is made. Similarly, the time-evolving module just forwards its previous evolving user factors to the next time period to provide an integrated recommendation.

We provide an illustrative example, where we randomly choose five test users with user ids: \{870391, 1197396, 757756, 920368, 1918714\} and remove their interactions in a given period (i.e., period = 4). We compare the prediction performance with the regular model where these interactions are not removed to show the impact of completely missing interactions in the current period. Table~\ref{no_interaction} shows that the proposed model provides quite robust results. In Period 4, due to missing interactions, the meta-learning module only utilizes global user factors. Similarly, for Period 5, the time-evolving module doesn't have inputs in Period 4 and hence forwards the user's time-evolving factors from Period 3. The performance with no interactions is slightly worse than having interactions.

\begin{table}[H]
\small
 \caption{Results with/without interactions in current period}
 \vspace{-2mm}
\centering
    \begin{tabular}{|r|r|r|r|r|}
\hline
\multicolumn{1}{|c|}{\bf Period} & \multicolumn{2}{|c|}{\bf Interactions}  & \multicolumn{2}{|c|}{\bf  No Interactions} \\
\cline{2-5}
\multicolumn{1}{|c|}{} &
 \multicolumn{1}{|c|}{\bf  RMSE} &  
 \multicolumn{1}{|c|}{\bf NDCG}
 &
 \multicolumn{1}{|c|}{\bf  RMSE} &  
 \multicolumn{1}{|c|}{\bf NDCG} \\
\hline 
4 &0.8720$\pm$0.04&0.3382 &0.8953$\pm$0.03 &0.3226\\

5 & 0.8466$\pm$0.04&0.3448 &0.8569$\pm$0.04&0.3415 \\
\hline
\end{tabular}
\label{no_interaction}
\end{table}

\subsection{Hyperparameter Tuning}
\begin{figure}
\centering
\begin{subfigure}{.245\textwidth}
  \centering
  \includegraphics[width=0.99\linewidth]{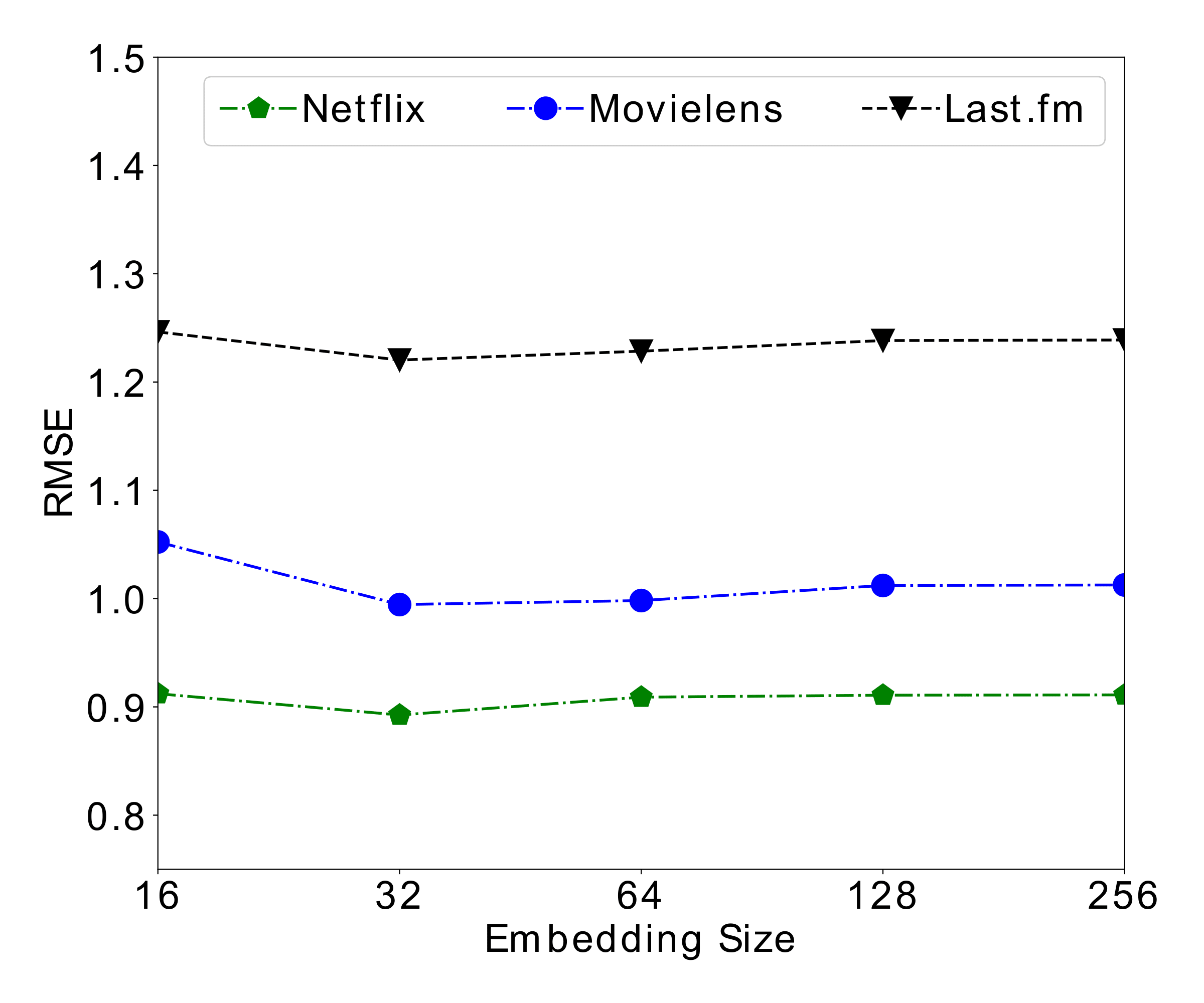}
  \vspace{-4mm}
  \caption{RMSE}
  \label{embed_rmse}
\end{subfigure}%
\begin{subfigure}{0.245\textwidth}
  \centering
  \includegraphics[width=0.99\linewidth]{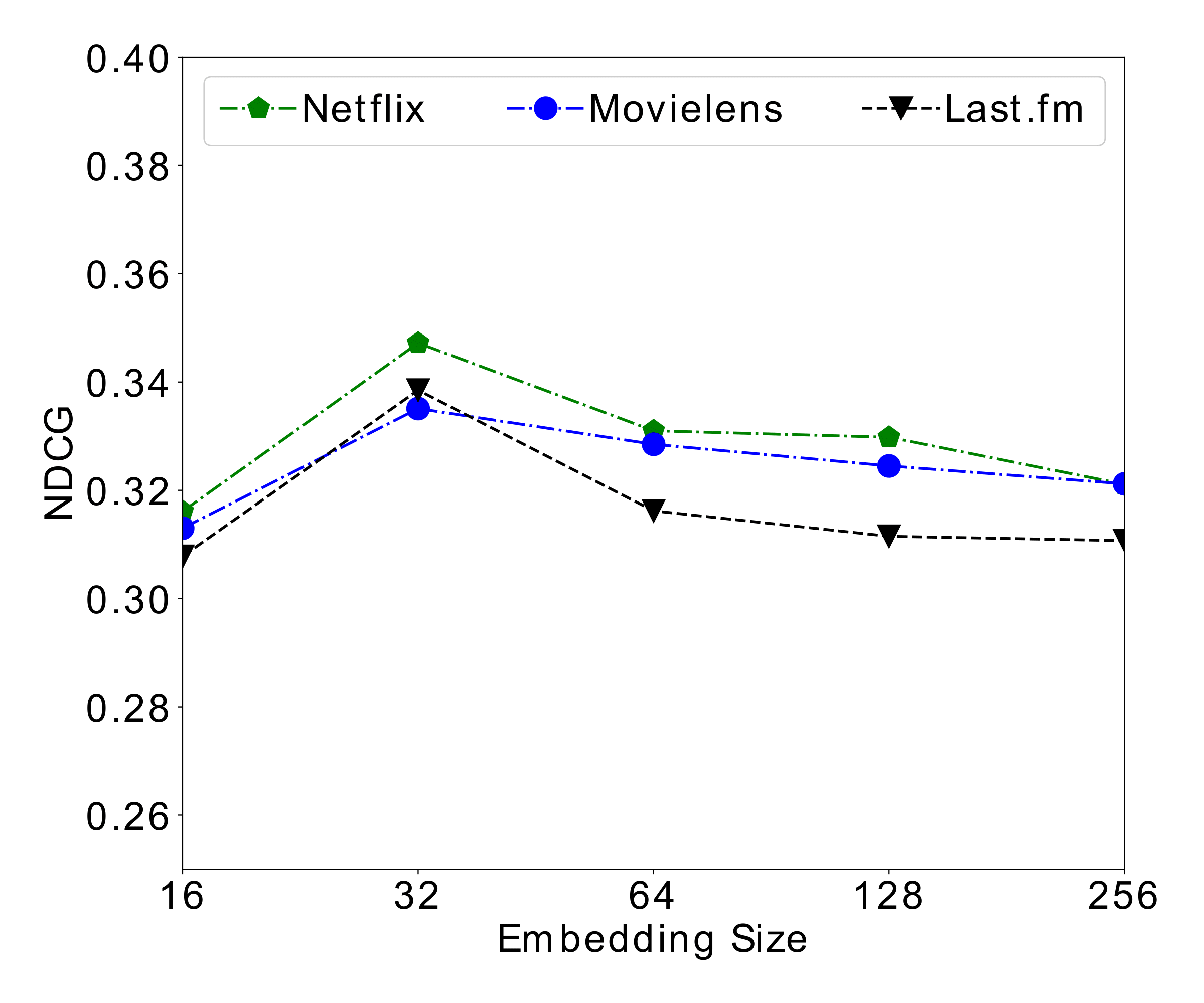}
  \vspace{-4mm}
  \caption{NDCG}
  \label{embed_ndcg}
\end{subfigure}
\vspace{-4mm}
\caption{Impact of item embedding size
}
\label{fig_emb}
\vspace{-4mm}
\end{figure}

We perform fine-tuning of model learning rates through a grid search from \{1e-5, 1e-4, 1e-3, 1e-2, 1e-1\} and select best value as $\alpha = \beta = \gamma = 1e^{-4}$. In our implementation, we learn user embedding from the model but item embedding is generated from the corresponding attributes. We embed each attribute (or feature) of an item in a vector of size 32. Each item in movies datasets has 5 attributes (genre, director, actors, plot, and overall rating), which are concatenated to generate the final item embedding of size 160. We applied different vector sizes [16,32,64,128,256] for each attribute and at 32 we got the optimal performance. Increasing the size didn't improve the results as shown in Figure ~\ref{fig_emb} for three datasets. So, we choose 32 for each attribute and the final embedding size is set as 160.

\section{Periodical Recommendation Results}
We report the detailed result for each prediction period for all three datasets to demonstrate how the predictions evolve over time. 

\paragraph{Netflix.} 
As can be seen from Table~\ref{netflix_period}, in most periods, the proposed model performs better than others except for a few periods like 3 and 6, where the proposed model slightly under-performs the meta-learning model. A possible explanation is that in these periods, time-specific user interest might largely deviate from the time-evolving user factors, and hence their combined recommendation is less accurate.

\begin{table*}[!t]
    \centering
    \caption{Netflix Periodic RMSE Results}
\label{netflix_period}
\resizebox{0.8\textwidth}{!}{
    \begin{tabular}{|p{0.9cm}|p{0.9cm}|p{1.6cm}|p{0.9cm}|p{0.9cm}|p{1.1cm}|p{2cm}|p{1.5cm}|p{0.9cm}|p{0.9cm}|p{1.5cm}|p{1.5cm}|}
    \hline
    \textbf{Period}&\textbf{SVD++}&\textbf{timeSVD++}&\textbf{CKF}&\textbf{MeLU}&\textbf{DeepFM}&\textbf{Wide \& Deep}&\textbf{GC-MC}&\textbf{Caser}& \textbf{DIEN}&\textbf{ML-ICS}&\textbf{Proposed}\\
    \hline
      1 &0.9813 & 0.9840&0.9552 &0.9201   &1.2706 &1.3178 & 1.2912& 1.2845& 1.3102& 0.9335&0.9205\\
      2 &0.9862  & 0.9895& 0.9683 &0.9481 & 1.1117&1.1321 & 1.2903& 1.2882& 1.3033&0.9498 &0.9258\\
      3 &  0.9795& 0.9795 & 0.9693&0.9440  & 1.0489&1.0921 & 1.2732& 1.2634& 1.2724&0.9445&0.9479\\
      4  &0.9742 &0.9708 & 0.9524&0.9374 & 1.0896&1.0777 &1.2613&1.2404&1.2543 &0.9401 &0.8771\\
      5  & 0.9800&0.9710 & 0.9432& 0.9415  &1.0630 & 1.0115& 1.2311& 1.2127& 1.2167&0.9426&0.9279\\ 
      6  & 0.9757&0.9740 & 0.9564& 0.9495  & 0.9911& 0.9904& 1.1374& 1.1206& 1.1515& 0.9487& 0.9533\\
      7 &0.9744 &0.9725 &0.9612 & 0.9506 & 1.0323&0.9895 &1.1068&1.1745&1.0984 &0.9622&0.8723\\
      8  &0.9766 & 0.9661&0.9526 &0.9418  & 0.9706&1.0122 & 1.049& 1.1132& 1.0401&0.9517&0.8789  \\ 
      9  &0.9737 &0.9603 &0.9412 & 0.9424 &0.9902 &0.9897 &1.0442&1.0724&1.0311 &0.9503& 0.8786\\
      10 & 0.9726&0.9597 &0.9228 &0.9391  &0.9961 &0.9655 & 1.0918&1.0843&1.0511&0.9430& 0.9209\\
      11  & 0.9722&0.9595 & 0.9332&  0.9360  & 0.9497&1.0570 & 1.03403& 1.0563& 1.0615&0.9315 & 0.8997\\
      12  &0.9731 &0.9511 &0.9541 & 0.9438  &0.9761 &  0.9648&1.0885&1.0245&1.0528 &0.9396& 0.9427\\
      13  &0.9361 &0.9556 &0.9243& 0.9351  &0.9710 & 0.9540& 1.0451& 1.0373& 1.0417&0.9306&0.8746\\ 
      14  & 0.9658&0.9431 & 0.9416& 0.9457  & 0.9720&0.9635 & 1.031& 1.0143& 1.0531& 0.9487& 0.8824\\
      15  &0.9673 &0.9457 &0.9358 &0.9378  & 1.0050& 0.9598& 1.0162&1.0142&1.0237&0.9441&0.8812\\
      16  & 0.9670& 0.9340&0.9337 &0.9428  &0.9866 &0.9550 & 1.0254& 1.0078& 1.0145&0.9468&0.8340 \\   
      \hline
\end{tabular}}
\end{table*}

\begin{table*}[!htpb]

    \centering
    \caption{Last.fm periodic RMSE results}
\label{last_period}

\resizebox{0.8\textwidth}{!}{
\small
    \begin{tabular}{|p{0.9cm}|p{0.9cm}|p{1.6cm}|p{0.9cm}|p{0.9cm}|p{0.9cm}|p{1.1cm}|p{2cm}|p{0.9cm}|p{0.9cm}|p{1.4cm}|p{1.4cm}|}
    \hline
    \textbf{Period}&\textbf{SVD++}&\textbf{timeSVD++}&\textbf{CKF}&\textbf{DPF}&\textbf{MeLU}&\textbf{DeepFM}&\textbf{Wide \& Deep}& \textbf{Caser}& \textbf{DIEN}&\textbf{ML-ICS}&\textbf{Proposed}\\
    \hline
      
1 & 1.6507& 1.6920&1.8353 &1.6953 &1.3153 &2.1434 &2.1993 &1.8132& 2.4720&1.3233& 1.3173\\
      
      2 &2.3052 &2.1647 &1.3964 &1.3238 & 1.5567&2.1336 &2.1675 &1.3854&2.1291 &1.5407&1.5320\\
      
      3 & 2.1972&1.9160 & 1.7007&2.0814 & 1.3408&2.0674 &2.1194 & 1.6772 &2.1065&1.3531&1.3465\\
      
      4 &2.1574 &1.7385 & 1.5462& 1.4654&1.1346&2.0757 & 2.0464& 1.5232&1.9520 &1.1287&1.1047\\
      
      5 & 1.5858&1.6520 & 1.6602& 1.5214&1.1687& 2.0202&2.0033 & 1.5843&1.9876 &1.1732& 1.1578\\
      
      6  &1.7879 & 1.8460& 1.5555& 1.3815& 1.2602& 2.0763& 2.1651& 1.5278& 1.8816&1.2821&1.0809\\
      
      7 &1.5348 & 1.6498& 1.6527& 1.3448& 1.6359& 1.9372&1.9595& 1.6227 & 1.8392&1.6324&1.6226\\
      
      8  &1.7527 & 1.7350& 1.5798& 1.3617&1.5987&1.8883 &1.8839& 1.5482 &1.8806 &1.6037 &1.4270\\
      
      9 &2.2419 &1.9977 &1.4311 &1.5085 & 1.4493& 1.9546& 2.0015 &1.4326& 1.9011& 1.4488&1.3338\\
      
      10 &1.9101 & 1.7901& 1.4572& 2.3051&1.1753 & 1.8346&1.8766 & 1.4104& 1.8743&1.1714&1.0368\\
      
      11 &1.5709 & 1.4704& 1.3067& 1.4337& 1.1528&1.7557 &1.8178 &1.2974& 1.8515 &1.1553&1.0864\\
      
      12 &1.5864 & 1.5760& 1.3353& 1.2409&1.2226 &1.7666 &1.7346 & 1.3136& 1.8483&1.2105&1.0933\\
      
      13 &1.5228 & 1.3637& 1.4613& 1.0546& 1.0546&1.7518 & 1.7192& 1.4463& 1.8617&1.0720&1.0123\\
      
      14 &1.7330 & 1.5966& 1.6130& 1.2670&1.0809 & 1.6417& 1.6800& 1.5812&1.8445 &1.0912&1.0603\\
      
      15 & 1.9891& 1.6644& 1.4429&1.3264 & 1.1094&1.6815 & 1.7253& 1.4293&1.7623 &1.0996&1.0824\\
      
      16 &1.4638 & 1.4106& 1.4334& 1.2547& 1.3206& 1.6900&1.9963 & 1.4017& 1.7456&1.3322&1.0688\\ 
      
      \hline
\end{tabular}}
\end{table*}

\begin{table*}[h!]

    \centering
    \caption{Movielens periodic RMSE results}
    \small
\resizebox{0.8\textwidth}{!} {
    \begin{tabular}{|p{0.9cm}|p{0.9cm}|p{1.6cm}|p{0.9cm}|p{0.9cm}|p{1.1cm}|p{2cm}|p{1.5cm}|p{0.9cm}|p{0.9cm}|p{1.5cm}|p{1.5cm}|}
    \hline
    \textbf{Period}&\textbf{SVD++}&\textbf{timeSVD++}&\textbf{CKF}&\textbf{MeLU}&\textbf{DeepFM}&\textbf{Wide \& Deep}& \textbf{GC-MC}& \textbf{Caser}& \textbf{DIEN}&\textbf{ML-ICS}&\textbf{Proposed}\\
    \hline
      1 &1.0615 & 1.0496&1.1155 &1.2234&1.5341&1.3719& 2.3345& 2.2941& 2.3438&1.2412&1.2253\\
      2 &0.9954&0.9932&0.9847&1.1613&1.3659&1.2686& 1.7912& 1.7247& 1.7065&1.1803 &1.0444\\
      3 & 1.0445 &0.9982&1.0305	&0.9341&1.2267&1.2585&1.2576&1.2289&1.2019 &0.9423&0.8487\\
      4  &1.1499	&1.1189&1.0508	&0.9776&1.2492&1.2346& 1.1332& 1.1223& 1.1623&0.9711&0.9084\\
      5  & 1.0918&	1.0611&	1.1468&1.0308&1.1615&1.1052&1.1346&1.1286&1.1587 &1.0514&0.9053\\ 
      6  & 1.0773&	1.0688&	1.0699&1.1377&1.0994&1.0672&1.1112&1.1021&1.1421 &1.1127&1.0377\\
      \hline
\end{tabular}}
\label{movielens_result}
\end{table*}

\paragraph{Last.fm Datasets.}
The period-wise results for the Last.fm dataset 
on one run is shown below in Table~\ref{last_period}. The proposed model achieves good results in this implicit feedback (i.e., counts) dataset. The high variation of the counts indicates users' music listening habits are fluctuating significantly. The results in Table~\ref{last_period} show that matrix factorization and deep learning baseline models are less effective in capturing those variations. In contrast, the proposed model simultaneously captures those variations in the form of users' specific biases and the gradual shift of preferences effectively.

\paragraph{Movielens Datasets:} Periodic results for the Movielens dataset are shown in Table~\ref{movielens_result}. In the first period, we notice that meta-learning models are not performing well, whereas SVD models are performing well. This is because the dataset has very dense interactions in the first period. Meta-learning models only use $k$-shot for learning, but SVD models benefit from maximum interactions. For the proposed model, time-evolving user factors don't contribute in the first period. Also, time-specific factors are based on meta-learning. Hence, its performance is not better than the baselines, but the proposed model achieves a better performance in the subsequent periods.

\section{Time Complexity}
Denote the computational complexity of embedding, time-specific and time-evolving module for one round of backpropagation as $O(m_1)$, $O(m_2)$, $O(m_3)$. In one epoch, the model iterates through $T$ tasks, and in each task, the model performs a local update for the time-specific module on the support set of size $S$ and also updates the time-evolving module on both support $S$ and query set $Q$. After iterating through the tasks, a meta update is performed utilizing query set $Q$ on the time-specific module. Overall, the total time complexity is $O(m_1 + T(m_2 + m_3) (S+Q))$ for each iteration.
We also provide the actual time taken by the proposed model to complete one training iteration utilizing GeForce RTX 2070 GPU. The proposed model considers each user as a task where the time-specific module and time-evolving module take 0.0065s and 0.0023s, respectively, to train on one task. Considering Movielens-1M dataset, we used 80\% training users and hence total training time for both modules in each iteration are 31.40s and 11.11s. Similarly, the embedding module takes 3.14s. This gives the total time of 41.65s/iteration.

\section{Link for the Source Codes}

\url{https://github.com/ritmininglab/A-Dynamic-Meta-Learning-Model-for-Time-Sensitive-Cold-Start-Recommendations}

\end{document}